 \documentclass[final,5p,times,twocolumn]{elsarticle}

\usepackage{amssymb}
\usepackage{graphicx}
\usepackage{color} 
\newcommand{\bea}{\begin{eqnarray}}
\newcommand{\eea}{\end{eqnarray}}
\newcommand{\be}{\begin{equation}}
\newcommand{\ee}{\end{equation}}

\journal{Physica C}
\begin{document} 
\begin{frontmatter}
\title{
Phase transitions to dipolar clusters and charge density waves in high T$_c$ superconductors}
\author[1]{M.\ Saarela}\author[2]{and F.\ V.\ Kusmartsev}
\address[1]{Department of Physics, P.O. Box 3000, FIN-90014, University of Oulu, Finland}
\address[2]{Department of Physics, Loughborough University, LE11 3TU, UK}

\begin{abstract}
We show that doping of hole charge carriers leads to formation of electric dipolar clusters in cuprates. They are  created by many-body interactions between the dopant ion outside and holes inside the CuO planes.  Because of the two-fold degeneracy holes in the CuO plane cluster into  four-particles resonance valence bond plaquettes bound with dopant ions.  Such dipoles may order into charge-density waves (CDW)  or stripes or form a disordered state  depending on doping and temperature. The lowest energy of the ordered system corresponds to a local anti-ferroelectric ordering. The mobility of individual disordered dipoles is very low at low temperatures and they prefer first to bind into dipole-dipole pairs.  Electromagnetic radiation interacts strongly with electric dipoles and when the sample is subjected to it the mobility changes significantly. This leads to a fractal growth of dipolar clusters. The existence of electric dipoles  and CDW induce two phase transitions with increasing temperature, melting of the ordered state and disappearance  of the dipolar state. Ferroelectricity at low doping is a natural consequence  of such dipole moments. We develop a theory based on two-level systems and dipole-dipole interaction to explain the behavior of the polarization as a function of temperature and electric field.
\end{abstract}
\begin{keyword}
high Tc superconductors \sep
charge density wave\sep
pseudogap\sep
dipolar cluster

\PACS 74.72.Gh \sep 
74.72.Kf


\end{keyword}
\end{frontmatter}



\section{Introduction}\label{intro}
Since the discovery of the hole-doped cuprate superconductors a great deal of effort has been devoted to understand their complex behaviour. Recently  guided migration of doped oxygen atoms in La$_2$CuO$_{4+y}$ (LCO) 
has been successfully demonstrated with nano-scale synchrotron radiation scanning X-ray diffraction in the temperature range 180 K $<$T$<$ 330 K. \cite{FratiniNature,PocciaNature,Poccia25092012} Originally  randomly  distributed  dopant atoms in  the LaO plane  order into clusters  after many hours of illumination.   Moreover, these ordered nano grains grow and form a fractal structure  that extends through the sample. 
The cluster growth looks as if there is a continuous phase transition (or a critical state)  where there are
two competing long-range orders(LRO). There exists a potential barrier separating minima associated with these orders and it takes a time for the dopant oxygens to self-organize and reach the new stable ordered phase. Probably because of this  criticality the cluster sizes observed have a scale free distribution. 
The superconducting critical temperature depends on the type of the cluster order. There are two critical temperatures of the clusters LRO observed. It was found that the higher transition temperature is associated with the better order where the critical percolation through ordered oxygens takes place. These fundamental discoveries  \cite{FratiniNature,PocciaNature,Poccia25092012} pose questions:  Why does the radiation  increase mobility of oxygen ions and why do they order? Is this order critical in increasing the onset temperature to superconductivity?

The inhomogeneity of cuprate superconductors coupled with dopant atoms is well documented besides LCO also in other materials and in different experiments using scanning tunneling spectroscopy \cite{PhysRevB.54.13324-Roditchev,RoditchevEPL,Davis05,Zeljkovic20072012,Nature.Physics.LeBoeuf2013}, photoemission spectra (ARPES) \cite{PhysRevB.74.094512-Richard}  --   suggesting that the oxygen dopant-induced states are mixed with Cu atoms -- and measurements of dielectric constant \cite{PhysRevB.72.064513_Wang} etc.   

Very recently a surprising discovery of ferroelectricity also in La$_2$CuO$_{4+y}$ was reported at exceptionally low oxygen doping.\cite{PanagopoulosSciRep2015,Panagopoulos} To describe this effect the Ginsburg-Landau theory of the ferroelectricity, where the existence of the   magneto-electric coupling has been assumed, was  developed. \cite{PhysRevB.85.140405} Other models based on polaron fromation and vortex-antivortex pairs have been suggested. Here we propose  that it is more natural to assume that the ferroelectricity arises due to electric dipoles caused by the Coulomb interaction of the dopant ions.

Debate between two competing orders and preformed pairs of superconductivity in the pseudogap region focuses mainly in the interpretation of ARPES measurements on the nature of gap functions and Fermi arcs. \cite{Hashimoto_Natphys_2014,PhilMagKaminski,NatPhys2012Reber,Zhao29102013,NatMater_Hashimoto_2015,KondoNatComm2015} Very recent experiments found evidences of sharp phase transitions at temperatures between the superconducting critical temperature T$_c$ and the pseudogap temperature at T*. \cite{WuNatComm2015,Fujita09052014,ShekhterNature2013,He25032011} At zero temperature inside the superconducting dome two phase transitions are discovered by studying the behavior of the gap function as a function of doping. \cite{Vishik06112012} Also the temperature dependence of resistivity changes from quadratic to linear behavior inside the pseudogap region with increasing doping and temperature. \cite{Cooper2009,PNASBarisic} Optically stimulated ultrafast changes in the charge-density wave correlations  have been studied by femtosecond resonant x-ray diffraction indicating that charge  ordering and superconductivity are competing orders.\cite{PhysRevB.90.184514} However, in spite of many attempts there is still no clear physical mechanism of the pseudogap and why there are phase transition between the pseudogap temperature T$^*$ and T$_c$.

Important experimental findings for our model come from a peculiar temperature dependence of the Hall coefficient, R$_H$,  measured in Sr doped La$_{2-x}$Sr$_x$CuO$_4$ (LSCO) \cite{PhysRevLett.92.197001,PhysRevB.75.024515-Ono}. These results have been analyzed within a two-band model \cite{PhysRevLett.97.247003-Gorkov}. It was argued that the charge carrier density, in that case the hole density, can be written as the sum of  two components
\be
n_{\rm h}(x,T)=n_0(x) +n_1e^{-\Delta(x)/2k_BT}\,.
\label{Gorkov}
\ee
The first term depends on doping $x$ and is independent of temperature $T$. The second term is of activation type contribution with a doping dependent activation energy $\Delta(x)$ multiplied by a constant $n_1$. 
The analysis \cite{PhysRevLett.97.247003-Gorkov} suggested that at small doping, $0.01<x<0.08$, and below the room temperature each dopant atom creates one hole and hence $n_0(x)=x$. At higher temperatures the hole density increases rapidly and for each $x$ that temperature behavior was well fitted with the activation type exponential component in Eq. (\ref{Gorkov}). The constant value $n_1=2.8$  indicated that more than one hole was activated. Even larger value $n_1\approx 4$ was found by Ono et al. \cite{PhysRevB.75.024515-Ono}.  The true nature of this two-component behaviour  and the activation energy $\Delta(x)$ was left as an open question. At higher doping $0.08<x<0.21$ and below the room temperature the hole density increases as a function of doping much faster than the number of dopant atoms and, more strangely, for a given $x$ when the temperature increases from 0K to 50K the  carrier density decreases. This suggests that the charge density can fluctuate between the two bands. 
\begin{figure}[t]
\begin{center}
\includegraphics[width=0.3\textwidth]{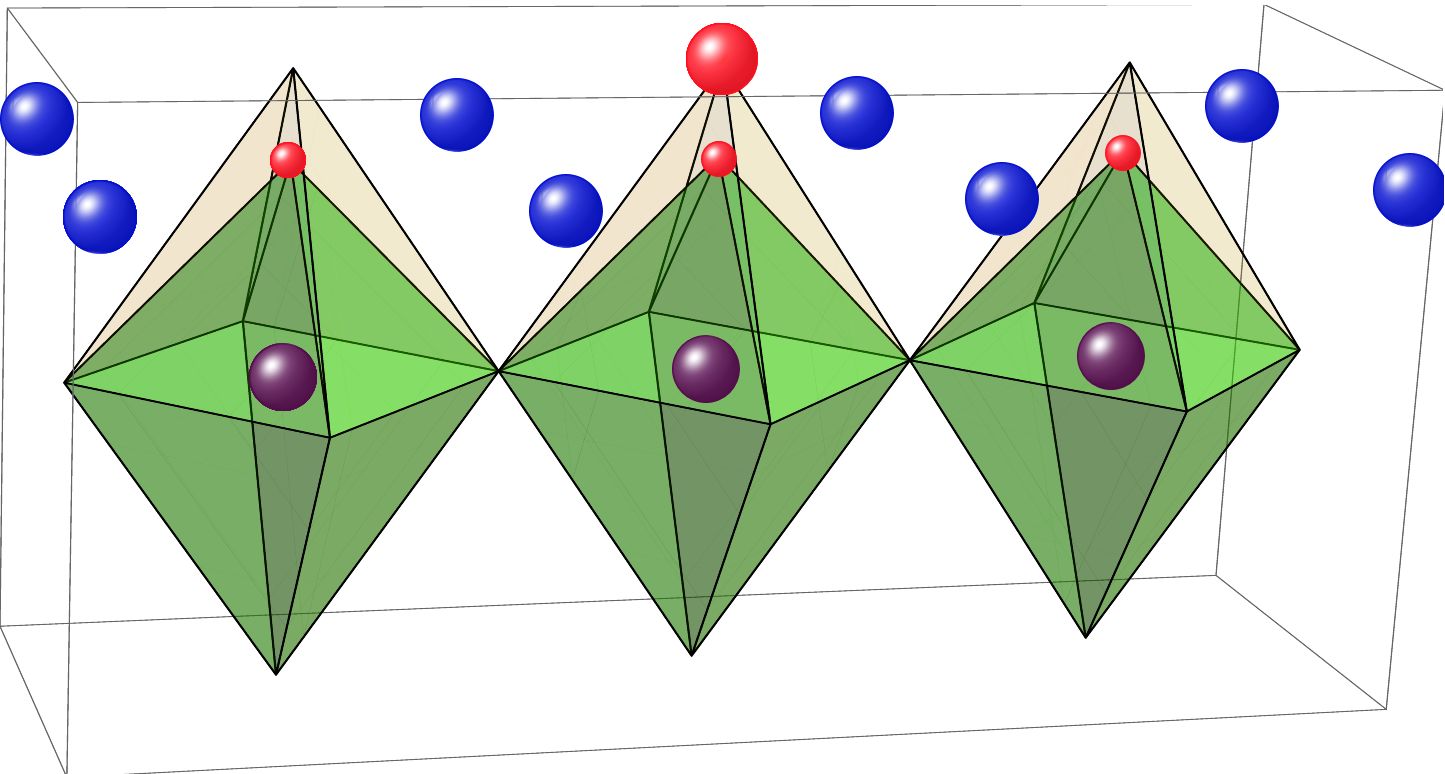}
\includegraphics[width=0.3\textwidth]{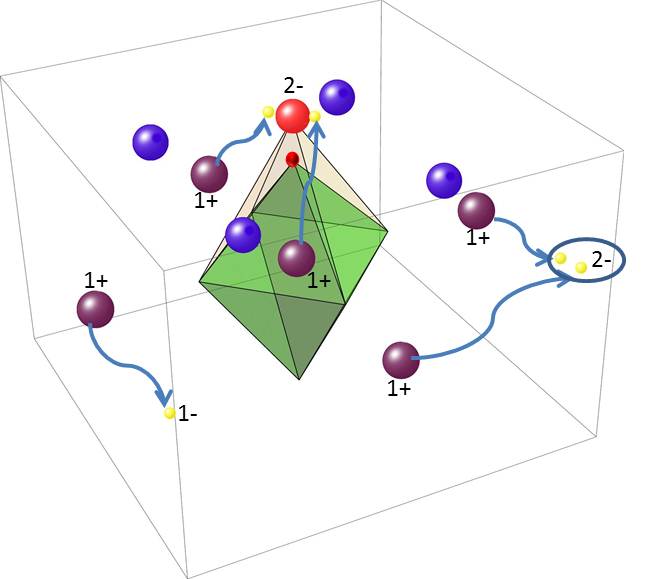}
\end{center}
\caption{Parts of the La$_2$CuO$_{4}$ crystal structure with one dopant impurity (topmost red sphere) are shown in two panels. Green octahedra have oxygens in all corners and Cu atom at the center (purple sphere). The apex oxygens (red spheres) and La atoms (blue spheres) form one of the two (LaO)$^{1+}$ spacer layers, which separates the (CuO$_2$)$^{2-}$ planes.  In the upper panel we show how  dopant impurity atom with its negative charge pushes the apex oxygen down from the top of light red octahedron to the top of the green octahedron. This restores the $e_g$ symmetry of the Cu orbitals (the anti-Jahn-Teller effect), which releases electrons from Cu atoms. 
 Note, that  there original ionic states  on Cu$^{2+}$ have  small positive charge transfer gap $\Delta_{CT}$. But the formal valence state Cu$^{3+}$, which should  be created by doping, 
have large negative $\Delta_{CT}<0$. This means that 
 the states should be rather represented as Cu$^{3+}\rightarrow$  Cu$^{2+}L$, (see, the Ref. \protect\cite{Zhang_Rice_Singlet} for a detail), where the doped holes would go not so much to the d-shells but rather to the oxygens, creating oxygen holes (though the quantum numbers of the respective states would be the same as those of Cu$^{3+}$).
In the lower panel we show schematically how electrons are redistributed. Thus, each Cu (precisely saying the Cu atom with surrounding oxygens) can "loose" effectively one electron.
Two of them are captured by the impurity oxygen to fill its p-shell.  This extra charge repels nearest neighbour Cu electrons and altogether five holes are revealed. The degeneracy allows four holes to bind into the ground state of the impurity  charge 2- and form a dipolar cluster.
}
\label{fig:LaCuO}
\end{figure}

In this paper we propose a microscopic approach \cite{KuSa2009} where bound states appear because dopant atoms 
( e. g. Sr$^{2+}$ are replacing the La$^{3+}$)  and therewith inducing  extra negative charges into spacer layers.  At low doping these negative charges are able to bind holes. The key ingredient in binding holes into dopant atom is the near degeneracy of the d-state bands with $x^2-y^2$ and $z^2$ symmetry in the CuO plane. The Jahn-Teller distortions in cuprates, such as La$_2$CuO$_4$, elongate the oxygen octahedron (green in Fig.1) surrounding the central Cu ion and split the degeneracy  of the associated $e_g$ orbitals of the Cu $d^9$-state. This happens in the highest partially occupied  $x^2-y^2$ - orbital, which together with the doubly occupied O $p_x, p_y$-orbitals form a strong covalent bonding.  Their straight hybridization gives rise to the bonding, non-bonding and half-filled anti-bonding bands and predicts a good metal, in sharp contrast with experiments finding a large charge gap. The failure of the band theory indicates a strong Coulomb interaction (in particular, on site Hubbard $U$), which may exceed well the
bandwidth of the tight-binding, anti-bonding band.


If we take the parent compound as La$_2$CuO$_4$ then two charge carriers (holes) put into the Cu $x^2-y^2$ orbital  would experience a large energy penalty, ($ U\sim 10 eV$).
Since on site Hubbard  $U$ is much larger than the energy separation between the Cu $x^2-y^2$ and O $p-$orbitals, which defines  the charge  transfer gap $\Delta_{CT} \sim 1-2 ~eV$, and this in turn is much larger than the hybridisation integral, $t_{dp}\sim 0.350~ meV$\cite{Tohayama-2000,Tohayama-2003,Tohayama-1998}, single electrons are localized on Cu sites on the Cu $x^2-y^2$ orbitals forming the Cu$^{2+}$ states. Their spins are anti-ferromagnetically aligned (via the super-exchange interaction that involves virtual hopping to the neighbouring O $p$ - orbitals) to create the anti-ferromagnetic Mott insulator\cite{Anderson-1997}.

 Since these original ionic states  on Cu$^{2+}$ have positive charge transfer gap $\Delta_{CT}$, which is small relatively to $U$, the formal valence state Cu$^{3+}$, which with doping would already have large negative $\Delta_{CT}<0$ should be rather represented by Zhang-Rice singlet , i.e.  as Cu$^{3+}\rightarrow$  Cu$^{2+}L$, (see, the Ref. \cite{Zhang_Rice_Singlet} for details). There doped holes would go not so much to the d-shells but rather to the oxygens, creating oxygen holes (though the quantum numbers of the respective states would be the same as those of Cu$^{3+}$). In  these Zhang-Rice singlets the hole is distributed  between the Cu$^{3+}$ and four neighbouring oxygens. Having this in mind  to make it short below we call these Zhang-Rice singlets simply saying as Cu$^{3+}$ states.

 In our  approach we consider this Mott state and take into account that cuprates, like La$_2$CuO$_4$, are also layered ionic crystals. The (CuO$_2$)$^{2-}$ layer is negatively charged and two spacer layers (La$_2$O$_2$)$^{2+}$ are positively charged. Oxygen atoms form an octahedral cell around Cu$^{2+}$ and the apex oxygen atoms containing two extra electrons filling the p-shell are in the LaO- layer as depicted in Fig. \ref{fig:LaCuO}. In the parent anti-ferromagnetic insulator compound the degeneracy of cubic $e_g$ states is removed by the Jahn-Teller effect, which elongates the oxygens octahedron. Copper, on the other hand, is a transition metal and it can easily give out one electron and transfer into  Cu$^{3+}\rightarrow$  Cu$^{2+}L$ sharing the electron with neighbouring octahedron oxygens and forming the Zhang-Rice singlet. 

Dopant atoms settle in the LaO spacer layer substituting La atoms like in the case of  Sr doping or intercalated like oxygen atoms. Oxygen atoms are small in size and thus mobile. Surprisingly  that these atoms  play an important role in the vortex trapping\cite{Rykov-1999}.  They easily fill their p-shell with two electrons  and become ionized. When this happens the negative impurity charge pushes the negatively charged apex oxygen down (see, Fig.1). This apex atom displacement  restores the $e_g$ symmetry and helps to remove one electron from neighbouring Cu-ions to form Zhang-Rice singlet (the anti-Jahn-Teller effect).
\footnote{
 Note that with this anti-Jahn-Teller effect associated with the squeezing of the oxygen octahedra, the hole state has a two fold degeneracy that is in addition of the conventional Kramers degeneracy.}  
At large enough doping holes created in this way in Cu sites form a charge carrying hole ("Zhang-Rice singlet") band  in the CuO plane.
Many complexes containing specific transition-metal central ions with special valency show this effect. \cite{Mueller-Nobel} 
Every new hole appearing during hole-doping  in the parent compound leads to strong frustration of the original antiferromagnetic state. This leads to rapid suppression of antiferromagnetism in La$_{2-x}$Sr$_x$CuO$_4$ even by small hole concentration. 

\section{ Charge Density Waves composed of resonance  plaquettes and electrical dipolar moments}
    
Four holes bound by dopant atoms due to correlations \cite{SaarelaJSNM} form resonance plaquettes in a manner similar to the resonance valence bond  described by Anderson\cite{Anderson-1997}. These plaquettes together with dopant atoms induce electrical dipolar moments. There may  also arise a second nearly flat band\cite{Kugel-2008}. 

As discussed above in hole doped cuprates dopant atoms are located in spacer layers between CuO planes and charge carriers are confined within CuO planes.  Dipolar clusters exist in a broad range of doping from deeply under-doped up to over-doped region. They disappear in over-doped region due to screening associated with the increasing hole density. 
In the region of the optimal doping plaquettes are decoupled from dopand atoms and become mobile. Then they are playing a key role in the Planckian dissipation observed in all cuprates\cite{Zaanen_2004}. 
The dipole-dipole interaction between clusters has a strong directional dependence.
It is attractive when parallel dipoles pointing to the same direction are on top of each other and repulsive when they are side by side.

Dopant oxygen atoms are mobile  although their mobility is very low at low temperatures. Then at temperatures below 80 K, at very low doping and random distribution of dopant atoms it is likely that dipoles remain in random, glassy order. 
Mobility rises with temperature  and when the sample is subjected to electromagnetic (X-ray) radiation.   
The radiation interacts strongly  with electric dipoles as well as doped oxygen impurities.  They can be excited by irradiation and  moved to new positions  associated with the minimum of the total energy.  That is how  irradiation increases the effective mobility of  dopant atoms and ordering of dipoles. As the result  they reorganize themselves  with the help of the dipole-dipole interaction and  after waiting long enough, as it was done in Refs. \cite{FratiniNature,PocciaNature}, a short-ranged order arises. 

The interaction between  dipolar clusters is short-ranged and varies as a function of separation distance with the $1/r^3$ tail. Clusters move towards the minimum energy configuration, but, because of low mobility, nucleation can start at different parts of the sample and lead to patches of different directional stripy order with a fractal structure like in classical dipolar systems.  The electron-phonon interaction will further enhance this ordering tendency leading to charge density waves\cite{Kusmartsev-1999,Kusmartsev-2000,Kusmartsev-2000E}. 

 Any isolated charge associated with an impurity or a charge fluctuation in LaO spacer layer is inducing the anti-Jahn-Teller shift of the apex O ion\cite{KuSa2009,SaarelaJSNM} and therewith creates a resonance plaquette in CuO plane.  To describe these resonance plaquettes we developed 
the many-body variational theory where four holes are trapped by the impurity \cite{KuSa2009,Saa317,EHLcluster,Krotscheckimpu}
 and derived a proper energetic description of resonance plaquette states, which reproduces experimental pseudogap temperature within the two-fluid model \cite{PhysRevLett.97.247003-Gorkov,PhysRevB.75.024515-Ono}.

\section{ Many-body theory of the  dipole moment formation}
In order to describe properties of a single dipole we need to solve a quantum many-body problem where holes interact via long-ranged Coulomb interaction among themselves and with the isolated charge associated with an impurity inducing the anti-Jahn-Teller shift of the apex O ion \cite{KuSa2009,SaarelaJSNM,KuSaIJMPB2015}.  In fact the many body interaction results in
 the bound state of the impurity in the LaO plane with holes in the CuO plane. We calculate below the binding energy of holes forming the dipole as a function of doping. To make a quantitative description of  such a state  it is sufficient to use a continuum approximation, where mobile charged holes  are confined to a CuO plane. Their effective mass is determined by the band structure of the material with the use of the effective mass or the (kp-) method \cite{Bir-1974}. For a simplicity, one may start with a three band Hubbard model\cite{Zhang_Rice_Singlet,Emery1987} where the effective holes mass, $m\sim t_{pd}^{-1}$.
 The charge neutrality is implemented by the inert background, called the jellium model\cite{KuSa2009}. Holes are allowed in general to move in a 2D-plane and the impurity is set at a given distance $c$ away from the plane. Thus the impurity serves as a nucleation center of the dipole moment pointing to c-direction. 

The Hamiltonian of such a system is the sum of two terms, 
$
H=H_h+H_I
$
which includes the mutual Coulomb interaction and kinetic energy of holes, $H_h$ and the apex O ion interaction with holes, $H_I$,
\bea
H_h&=&-\sum_{i=1}^{N}\frac{\hbar^2}{2m}\nabla_i^2 
+\frac 1 2 \mathop{\sum^{N}_{{i,j=1},{i\ne j}}}
\frac{e^2}{4\pi\epsilon|{\bf r}_i-{\bf r}_j|}
\cr\cr
H_I&=&-\frac{\hbar^2}{2M}\nabla_0^2 
-\mathop{{\sum}}_{i=1}^{N}
\frac{2e^2}{4\pi\epsilon\sqrt{|{\bf r}_0-{\bf r}_i|^2+c^2}}\,.
\eea
The number of mobile holes is $N$, they have the mass $m$, charge $|e|$ and position ${\bf r}_i$. The impurity is placed at ${\bf r}_0$, its charge is $-2|e|$ and kinetic energy is controlled by the mass $M$. For a localized impurity we let the mass grow to infinity. The strength of the Coulomb interaction depends on the dielectric constant $\varepsilon$. It is convenient to use the atomic units where all distances are given in units of $r_0=r_s r_B$ and energies in Rydbergs. The parameter $r_s=1/\sqrt{\pi n_0} ~r_B$ is defined by the density $n_0$ of holes and the Bohr radius $r_B$. 

At low doping in the under-doped region the hole gas is dilute, but very strongly correlated.
 We assume that the hole band is two-fold degenerate, because the mechanism to produce the conducting hole band relies on the anti-Jahn-Teller type shift in almost degenerate energy levels. Under such circumstances, in strongly correlated regime, the difference between the fermionic and bosonic gases is of minor importance.  The cluster formation takes place in both cases although in slightly different regions of $r_s$ values \cite{EHLcluster,Krotscheckimpu}.
The bosonic part of the ground-state wave function which contains correlations between holes and impurities is chosen in the form of  the Jastrow-type variational ansatz\cite{FeenbergBook}
\bea
\Psi &=&e^{\frac1 2\sum^{N}_{i,j=1}
	u^{hh}(|{\bf r}_i,{\bf r}_j|)}
	\cr\cr 
\Psi_I&=&	e^{\frac1 2\sum^{N}_{i=1}
	u^{It}(|{\bf r}_i,{\bf r}_0|)} 
	e^{\frac1 2\sum^{N}_{i,j=1}
	u^{In}(|{\bf r}_i,{\bf r}_0|)}\Psi
\eea
 We have extended the conventional, many-body variational theory to the case where four holes are trapped by the impurity.\cite{KuSa2009} The wave function includes now the product of three components, hole-hole (hh) correlations, impurity-trapped holes (It) correlations and impurity non-trapped holes (In) correlations.
The correlation functions $u^{hh}(|{\bf r}_i,{\bf r}_j|))$, $u^{It}(|{\bf r}_i,{\bf r}_0|))$ and $u^{In}(|{\bf r}_i,{\bf r}_0|))$   are determined by minimizing the total energy $E$ of the hole gas and the chemical potential $\mu$ of the impurity. 
\bea
	E&=&	\frac{\langle\Psi| H_h | \Psi\rangle}{\langle\Psi| \Psi\rangle}
	\cr
	\mu 
	&=& \frac{\langle\Psi_I\vert H_I \vert \Psi_I\rangle}
		{\langle\Psi_I\vert\Psi_I\rangle}
	-\frac{\langle\Psi\vert H_h\vert \Psi\rangle} 
	{\langle\Psi\vert\Psi\rangle}- 4 E_{bin}\,,
\label{chemi-M}
\label{Euler}
\eea
where the last term is the sum of binding energies $4E_{bin}$ of four trapped holes. The maximum number of trapped holes  in the same state is determined by the Pauli principle. It is twice the degeneracy factor of the hole band, which we set equal to two. The fact that four holes are bound by the charge -2$|e|$ is caused by strong many-body correlations. As a result of the variational calculation we get the total and binding energies, chemical potential  and the hole-hole and hole-impurity distribution functions. For details we refer to Ref. \cite{KuSa2009}.

 \begin{figure}[tb]
\centering
\includegraphics[width=0.5\textwidth]{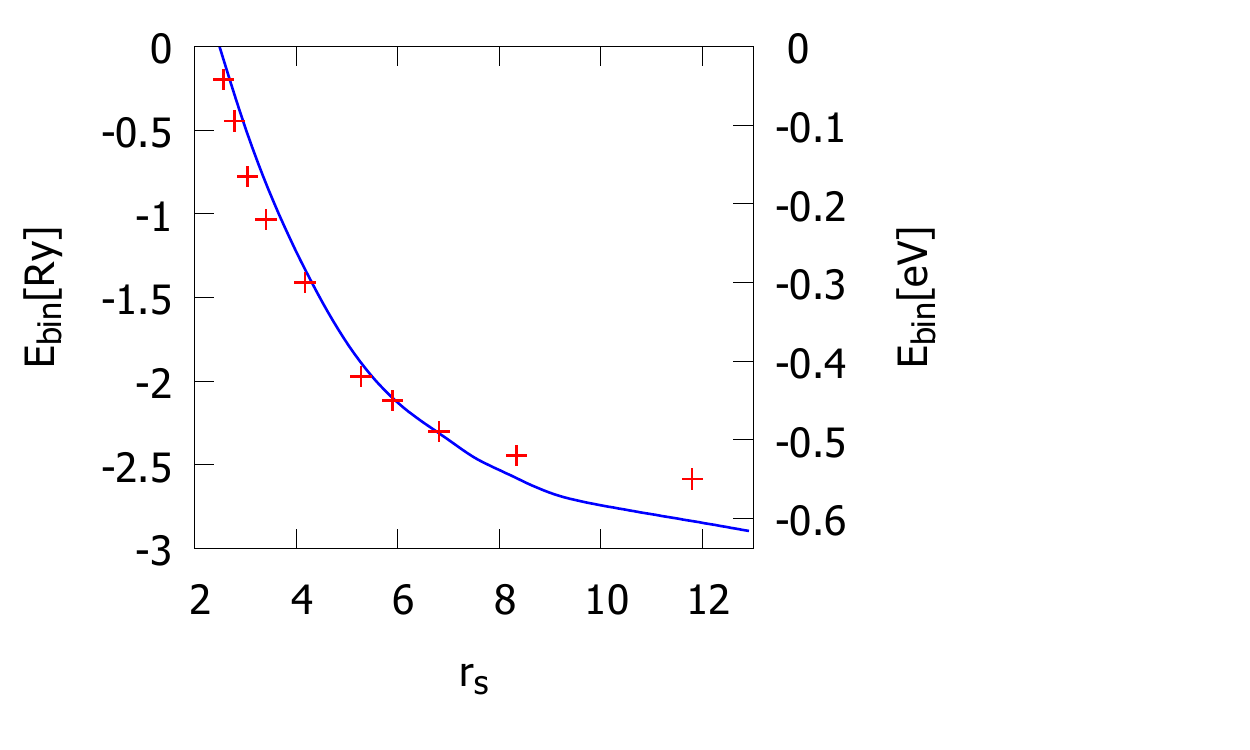}
\caption{The calculated binding energy of the hole cluster $E_{bin}(x)$ (solid blue curve) identified as the activation energy  and fitted to the Hall coefficient measurements \protect\cite{PhysRevB.75.024515-Ono,JPSJ.63.1441} (red plus signs). 
}
\label{activation}
\end{figure}

In Fig. \ref{activation} we show the binding energy $E_{bin}$ as a function of $r_s$ parameter. The binding energy vanishes when $r_s<2.5$ and saturates very slowly to the single hole limit  when $r_s\rightarrow\infty$. At zero temperature the number of holes in CuO layers is equal to $x$, the number of dopant atoms, but when the density is high enough, $r_s<2.5$, then four more holes per dopant atom are released. The maximum number of holes is one per Copper site and that is reached when $x=1/5=0.2$. This is clearly seen by Cooper et al. \cite{Cooper2009} as a change in the temperature dependence of the resistivity at x$\approx 0.19$, which is identified as a quantum critical point.  In the Gorkov-Teitelbaum two-fluid model\cite{PhysRevLett.97.247003-Gorkov} the activation energy $\Delta(x)$ vanishes also at $x\approx 0.2$. 

\begin{figure}[tb]
\center{
\includegraphics[width=0.5\textwidth]{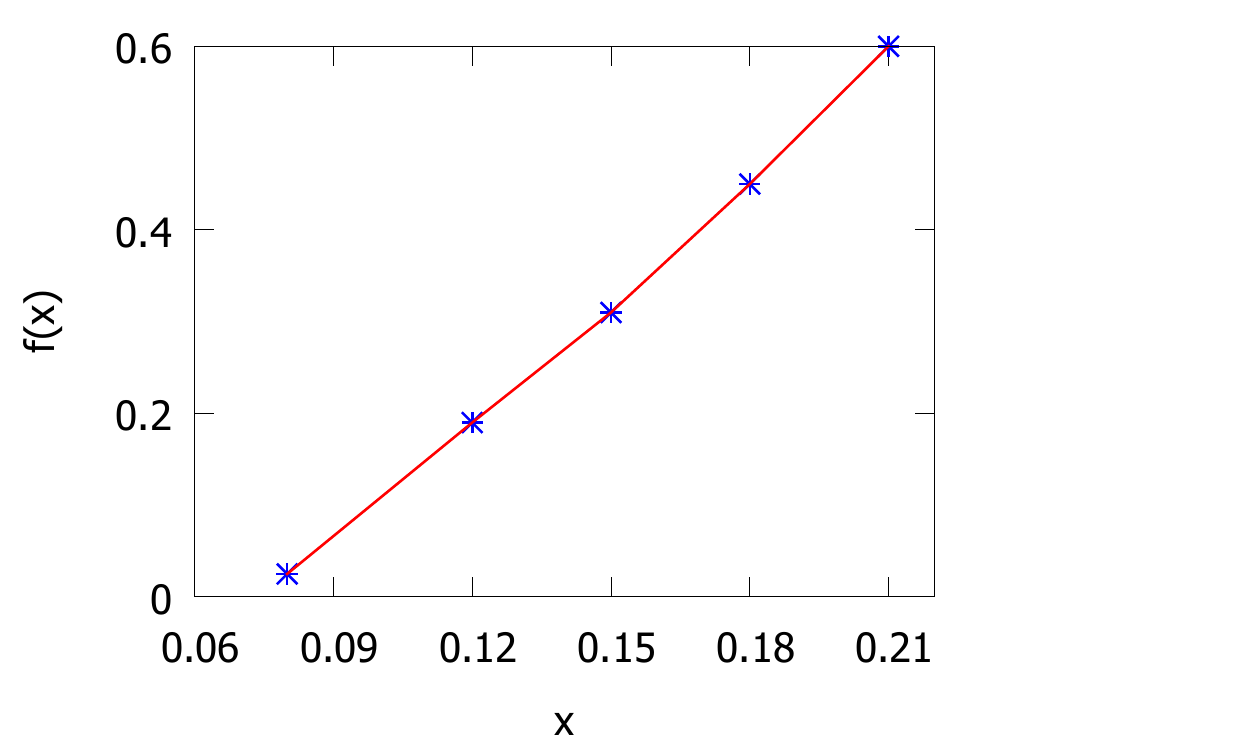}
}
\caption{The function $f(x)$ from Eq. (\ref{hole_density}) fitted to the Hall coefficient data by Ono el al. \cite{PhysRevB.75.024515-Ono} at values of doping show by blue stars. The solid line is just to guide the eye.}
\label{fluctuation}
\end{figure}

The cluster formation gives the microscopic foundation to the assumption of two kinds of charge carries, doped and activated. In order to allow the transfer of holes between these two components  seen in  the Hall coefficient data \cite{PhysRevB.75.024515-Ono} we have refitted  them with the formula,
\be
n_h=x(1+4*f(x))+4 ( 1-f(x))e^{-\Delta(x)/(2 k_B T)}
\label{hole_density}
\ee
The function $f(x)$ shown in Fig. \ref{fluctuation} gives the fraction of  holes freed from the clusters already at zero temperature. In an ideal system $f(x)$ should be a step function, zero at low doping and one when all holes are released from the trap. At zero temperature that happens at $x=0.2$ and the density of holes per Cu site $n_h=1$. That is the phase transition from dipolar cluster state to homogeneous metallic state. In practice high T$_c$ superconductors are inhomogeneous. There are high and low density regions and that is seen as a smooth behavior in the fitted $f(x)$. In the high density regions, $r_s<2.5$, the binding energy of clusters vanishes and all four bound holes per cluster are then free to carry the charge and out of the activation process. In the fit we have averaged the low temperature fluctuations between 0K$<T<$50K. In Fig. \ref{fluctuation} we show the fitted values. For low doping, $x<0.08$,  $f(x)$ vanishes and then increases up to 0.6 at the highest doping, x=0.21, measured. Taking strictly the zero temperature limit of experiments the values of $f(x)$ grow faster with increasing doping reaching unity when $x=0.2$, which indicates that all holes are freed. 

The fitted activation energies $\Delta(x)$  shown in Fig.  \ref{activation} by red symbols can be compared with our  binding energies $-E_{bin}(r_s)$ after the conversion of our Rydberg unit into electron volts and $r_s$ parameter scale into the  doping scale $x$.
In our notation the density of holes $n_h= 1/\pi (r_s r_B)^2=x/\pi r_{\rm Cu}^2$ where $r_{\rm Cu}\approx  1.9{\rm  \AA}$ 
is half of the separation distance between Copper atoms. We identify the point $r_s=2.5$ where $E_{bin}(x)$ vanishes with the value of doping x=0.22 which is slightly higher than the value 0.2 where the activation energy vanishes because of the fluctuations as pointed out above. 
That fixes the Bohr radius $r_B\approx 1.64{\rm \AA}$.  Then we are left with one parameter to convert the energy scale. By choosing 1Ry= 0.213 eV we get the fit shown in  Fig.  \ref{activation} by blue solid curve.   Taking into account these values of Rydberg and Bohr radius we get the dielectric constant $\epsilon=21$ and the effective hole mass $m= 6.7 m_e$ in units of the electron mass $m_e$. The dielectric constant found here is in reasonable agreement with the value  $\epsilon=29$ determined from Chen et al. experiments in LSCO \cite{Chen}. 

\begin{figure}[tb]
\center{
\includegraphics[width=0.5\textwidth]{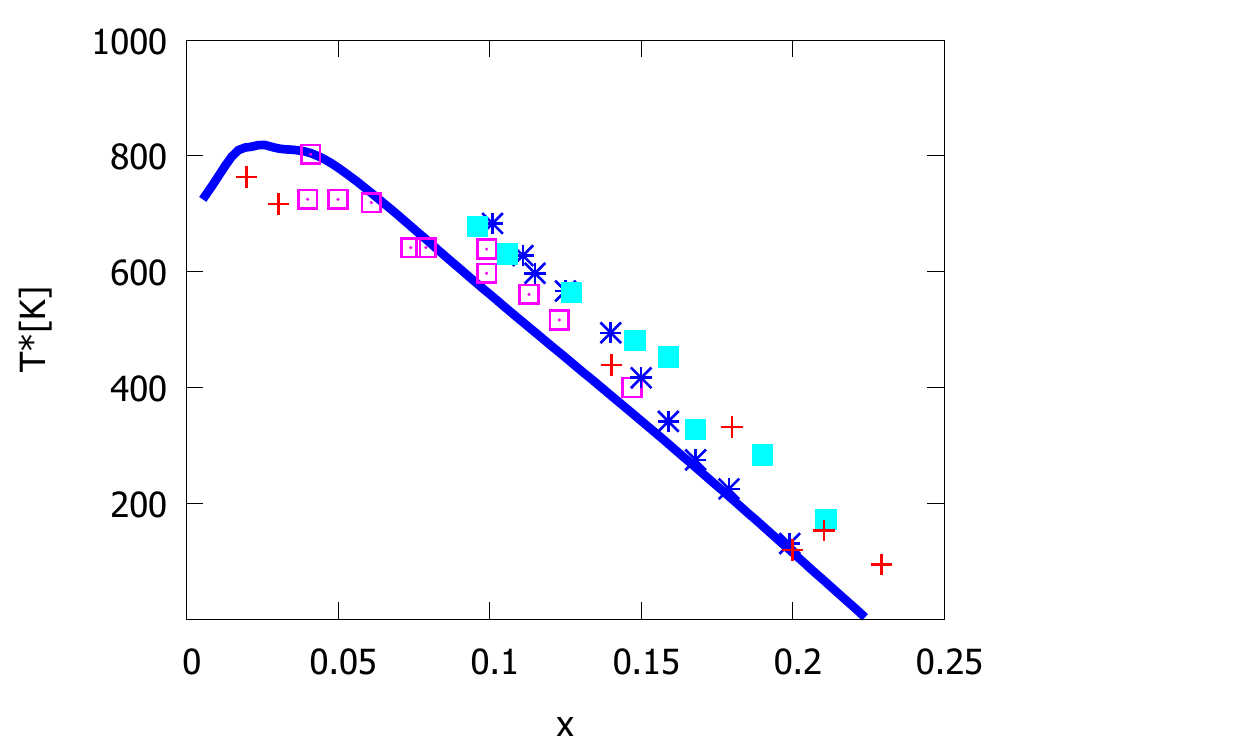}
}
\caption{The pseudogap temperature calculated from Eq. (\ref{pseudo}) is compared with the crossover temperature
of the resistivity curves (open squares)[\cite{Timusk_1999}],  the temperature of  the maximum magnetic susceptibility
(stars) [\cite{Yoshizaki_1990} ]  and (solid squares) [ \cite{Nakano_1994}], and with  the activation energy (plus-signs)[\cite{PhysRevLett.97.247003-Gorkov}].
}
\label{pseudogap}
\end{figure}

We identify the pseudogap temperature with the temperature when all trapped holes are released. Then 
\be
x=e^{-\Delta(x)/(2 k_B T^*)}\,,
\ee
and  the density of holes $n_h=5x$.  That is the phase transition line in temperature doping plane when dipolar clusters melt into the bad metal state. This transition is not sharp because of the density fluctuations, and one enters the Fermi liquid regime with regions of clusters. That would be some kind of an emulsion state. In Fig. \ref{pseudogap} we show the calculated pseudogap temperature
\be
T^*(x)=E_{\rm bin}(x)/(2 k_B \log(x)) 
\label{pseudo}
\ee
and find a good agreement with different experimental values. Doping dependence of  $T^*(x)$  is dominated by the binding energy because $\log(x)$ is a slowly varying function in the range $0.05<x<0.2$.

\begin{figure}[tb]
\center{
\includegraphics[width=0.5\textwidth]{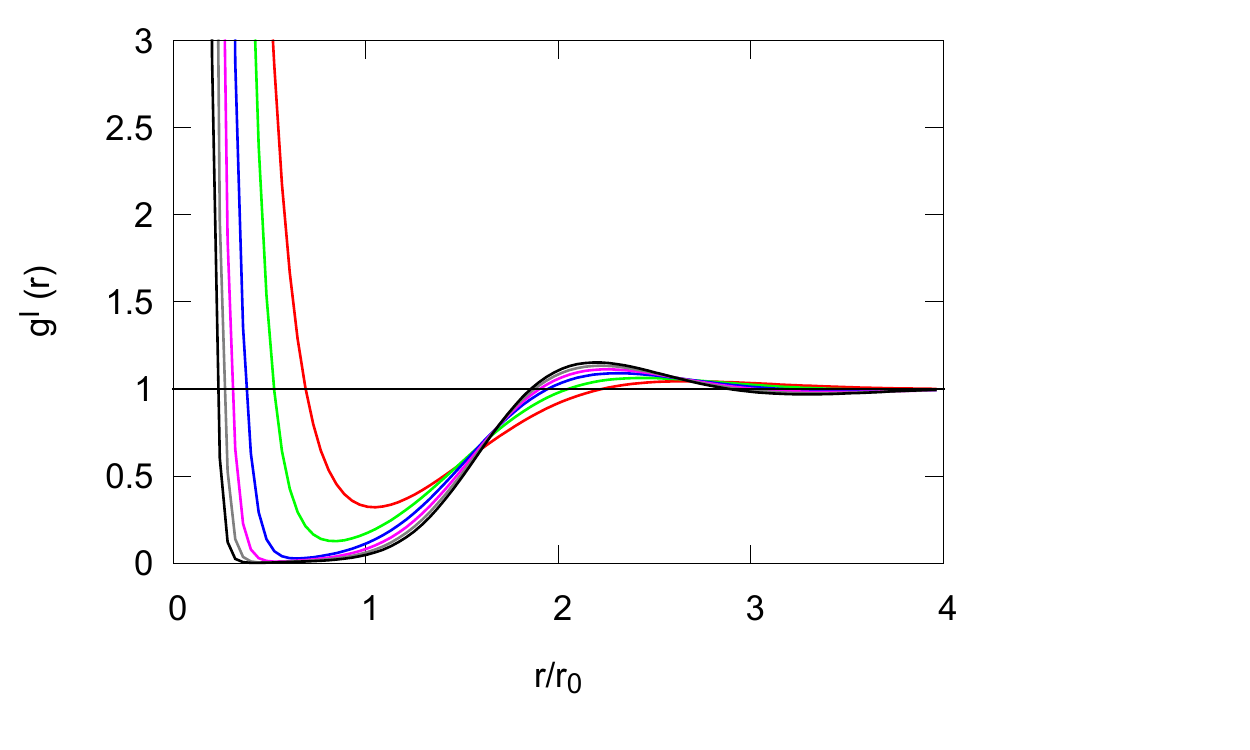}
}
\caption{The radial distribution functions of  holes in the CuO layer around the impurity charge $-2|e|$ in the LaO layer . The impurity is located at the origin and the peak around it is huge in this scale. The highest peak is at the lowest density when $r_s=12$ with the height 780. The curves from left to right are calculated at $r_s$=12, 10,8,6,4 and 2.5, respectively. All distributions are normalized to unity at $r\rightarrow \infty$. The deep minima at $r<1$ separates four bound holes from the continuum holes.}
\label{gbi}
\end{figure}

\begin{figure}[tb]
\center{
\includegraphics[width=0.4\textwidth]{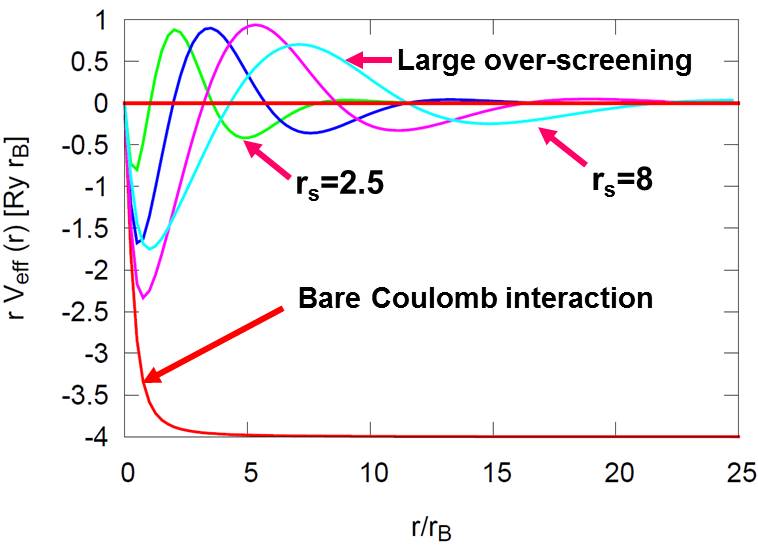}
}
\caption{The hole-impurity effective interaction  as a function of separation distance $r$ and multiplied by $r$ calculated at $r_s$=8,6,4 and 2.5 as marked in the figure. Also shown is the bare Coulomb interaction with the red curve at the bottom. }
\label{veff}
\end{figure}

\begin{figure}[tb]
\center{
\includegraphics[width=0.4\textwidth]{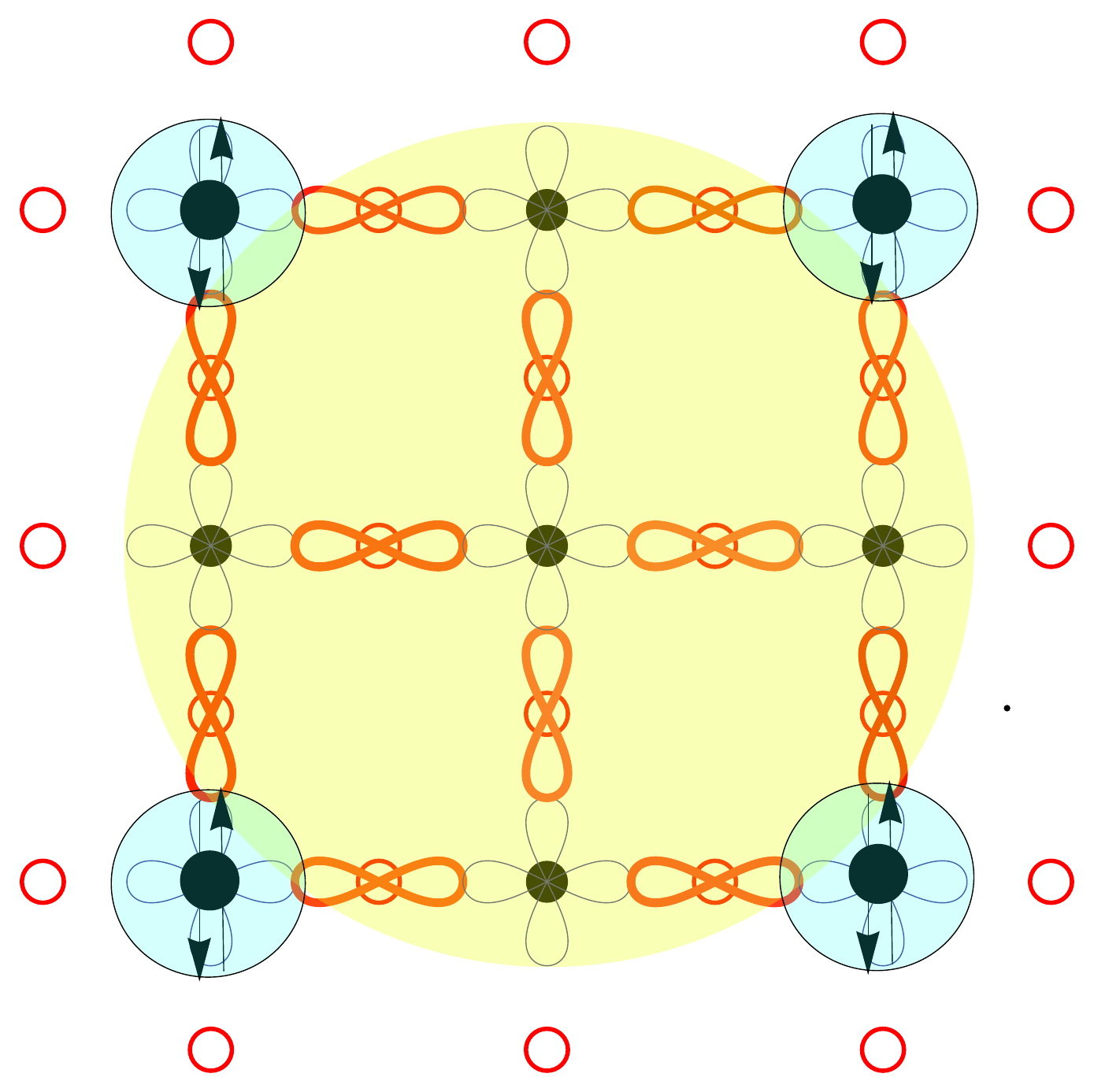}
}
\caption{A schematic  top view of the charged hole cluster in CuO plane of La$_2$CuO$_{4}$ noted by yellow disk. At the center of the cluster there is the Cu atom surrounded by four oxygen and other Cu atoms with the d and p-state orbitals shown respectively. Inside the cluster there are four holes trapped. Outside -  double occupied states (Mottness) arising to keep the electro-neutrality. 
}
\label{fig:topview}
\end{figure}

In Fig. \ref {gbi} we show  the radial distribution of  holes in CuO plane around an impurity, which is  located at  half a Bohr radius above the  CuO plane. The bound trapped holes correspond to a large peak in $g^I(r)$ around the origin. At the highest density where the activation energy vanishes ($r_s=2.5$, $r_0=4.1 {\rm\AA}$ and $x=0.22$) the radius of the bound state wave function barely covers the nearest neighbour Cu atoms, since the distance between Cu atoms is about 3.8{\rm \AA}.  When the hole density decreases  the range of wave function extends further covering the whole neighbouring octahedron, see, the Fig. \ref{fig:LaCuO} and the yellow circle in the Fig. \ref {fig:topview}. As pointed out earlier the Pauli principle allows only four particles to be bound in the lowest state and that is why one of the five covered holes (the maximum number of holes) is free to act as a charge carrier. That is consistent with the experimental fact that the number of holes is equal to the number of dopant atoms at low doping and the maximum hole density is reached when all five holes per dopant atom are released.

Finally in Fig. \ref{veff} we show the effective interaction between the impurity and holes. At short distances it is attractive, but many body effects strongly over-screen attraction at intermediate ranges leading to bound clusters.

Four bound, positively charged holes over-screen the impurity ion charge -2$|e|$. It means that two electrons are expelled from the octahedron as circled out  in Fig.  \ref{fig:LaCuO}. They are attracted to the positively charged LaO-layer and get loosely trapped by four apex oxygens squeezed to their symmetry positions in the same way as the impurity ion does. That is a new source of electron pairs, which will form a very flat band together with the bound holes in the CuO plane.  The binding energy of holes to the electron charges is roughly the same as to the impurity charge and that is why at low doping the factor in front of the activation contribution to the charge carriers is independent of $x$  in Eqs. (\ref{Gorkov}) and (\ref{hole_density}). 
This phenomenon is called Mottness. In Refs. \cite{RevModPhys.82.1719-Phillips,PhysRevB.80.132505-Chakraborty}, however,  a different microscopic mechanism for the Mottness  accounting for spectral weight transfer experiments has been  suggested.

The positively charged clusters in the CuO-plane expel surrounding holes further away and reveal the background electrons. In our jellium approximation this shows up in the hole distribution function $g^I(r)$ as a deep minimum around the central hole peak extending to 2$r_0$  (see, Fig.   \ref {gbi}).  At infinity the  distribution is normalized to one, which accounts for the fact that the charge of holes  compensates totally the background charge and the integral of $g^I(r)-1$ is equal to the impurity charge. 

%

\section{The dipolar clusters}

%
\begin{figure}[tb]
\center{
\includegraphics[width=0.4\textwidth]{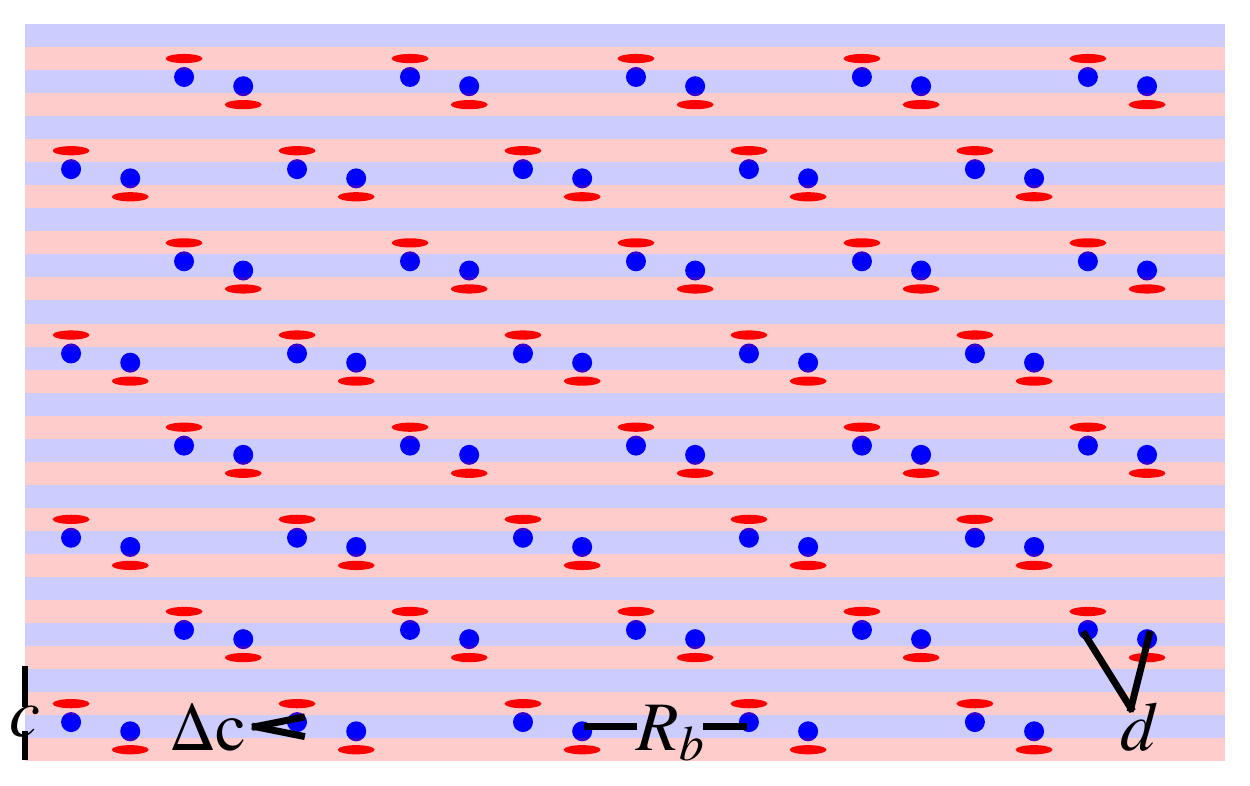}
}
\caption{Ordered structure of dipolar clusters in the layered superconductors in the bc-plane. Four-hole clusters (red) are in the CuO plane and i-O ions (blue) in LaO planes. The separation distance between  LaO and CuO layers is $c/4$. The  separation distance between bound dipoles with opposite orientation is $d$ and the distance between dipole-dipole pairs is $R_b$.
}
\label{fig:ldg}
\end{figure}

\begin{figure}[tb]
\center{
\includegraphics[width=0.4\textwidth]{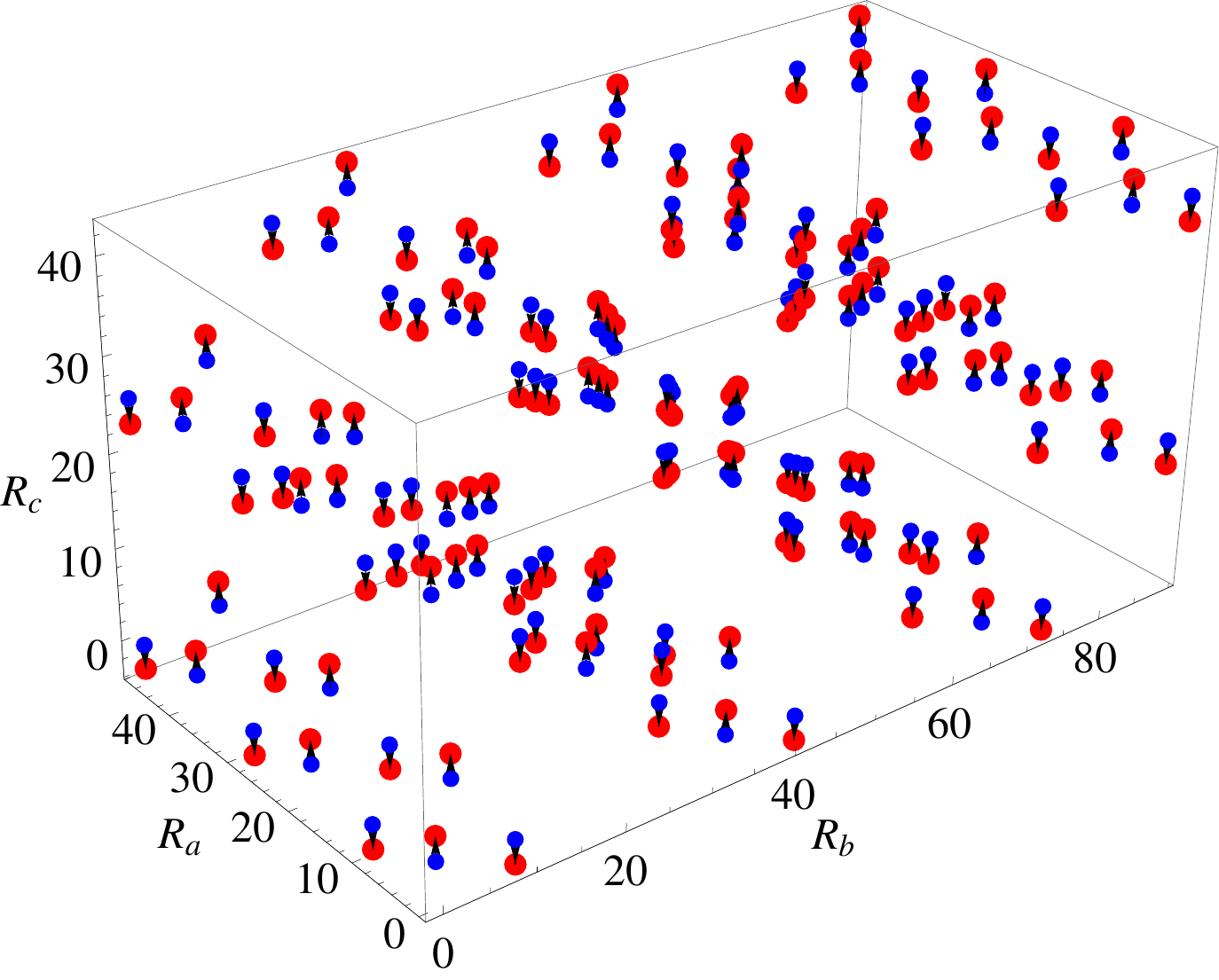}
}
\caption{Regularly ordered clusters in 3D. The red circles  determine the CuO plane. The i-O (blue circles) can be below or above that plane. The minimum energy in this configuration is reached when the dipole-dipole pairs are separated by $d$=13\AA.
}
\label{fig:3Ddg}
\end{figure}

The cluster of four holes in the CuO plane self-bound due to many-body effects to the excess  charge created in the LaO plane around the apex oxygen pushed by a dopant ion (the anti-Jahn-Teller effect) has a strong dipolar moment and low mobility. With increasing doping these clusters form a classical dipolar gas. The potential energy of $N$ dipoles is determined by the impurity charges $q_{\rm LaO}=-2|e|$ located in LaO layers, each of which creates the opposite sign charge cluster in a CuO layer $q_{\rm CuO}=2|e|$.  For simplicity, we assume that these charges in both layers are point like, heavy objects and calculate their Coulomb potential energy classically
\be
V=\frac{1}{4\pi\epsilon}\sum_{i\ne j}^N\sum_{\alpha,\beta}\frac{q_\alpha q_\beta}{|{\bf r}_\alpha(i)-{\bf r}_\beta(j)|}\,.
\ee
 It depends on the charges and dielectric constant $\epsilon$, but most importantly it depends on the orientation of clusters. The Greek indices refer to the location of the charge, either in LaO or CuO layer, and ${\bf r}_\alpha(i)$ to the position of the charge in that layer. In the calculated examples we have required the following regular order. A bound dipole pair is separated by a distance $d$ pointing to the $b$ direction in the $a,b$-plane. The pairs are then separated by distances $R_a$ and  $R_b$ in $a$ and $b$ directions, respectively.
The motion of holes is strictly two-dimensional in the CuO-layers, but dopant atoms  have some flexibility at high temperatures to diffuse between two LaO layers. The octahedron can then be polarized either from above or below as shown in Fig. \ref{fig:ldg}. With these assumptions the potential energy of two neighboring, oppositely oriented dipoles (d-d interaction) associated with the same LaO layer as a function of their separation distance $d$ is then
\be
V_{\rm dd}(d)=v(d,\Delta c)+v(d,c/2)-2v(d,c/4+\Delta c/2)
\ee
where $v(d,x)={4e^2}/{(4\pi \epsilon}{\sqrt{d^2+x^2})}$,  
$c=13.13{\rm  \AA}$ is the size of the unit cell in c-direction and $\Delta c\approx c/10=1.3{\rm  \AA}$ is the separation distance of the impurity charge centers in c-direction as shown in Fig. \ref{fig:ldg}. $V_{\rm dd}(d)$ has the minimum at $d=7.3{\rm  \AA}$ with the energy  -12 meV, when we use  the dielectric constant $\epsilon=21$ taken from our activation energy fit shown  in  Fig. \ref{activation}. This defines the temperature scale in the self organization of the ordered structures seen in the experiments \cite{FratiniNature,PocciaNature}.

The growth of nano grains under soft X-ray radiation \cite{FratiniNature,PocciaNature} takes place in the temperature range, 330K$>$T$>$180 K, well above the transition to superconducting state, but below the pseudogap where dipolar clusters exist.  The d-d interaction dominates the growth process. First appear bound d-d dimers in the a-b plane and they arrange themselves into anti-ferroelectric stripes as shown in Fig. \ref{fig:3Ddg}. The growth into larger clusters slows down the mobility and therefore patches of high density dipolar clusters can appear in different places of the sample. At high enough temperature depending on doping or hole density directional order melts and stripes disappear indicating a second phase transition inside the pseudogap region. 

\begin{figure}[tb]
\center{
\includegraphics[width=0.4\textwidth]{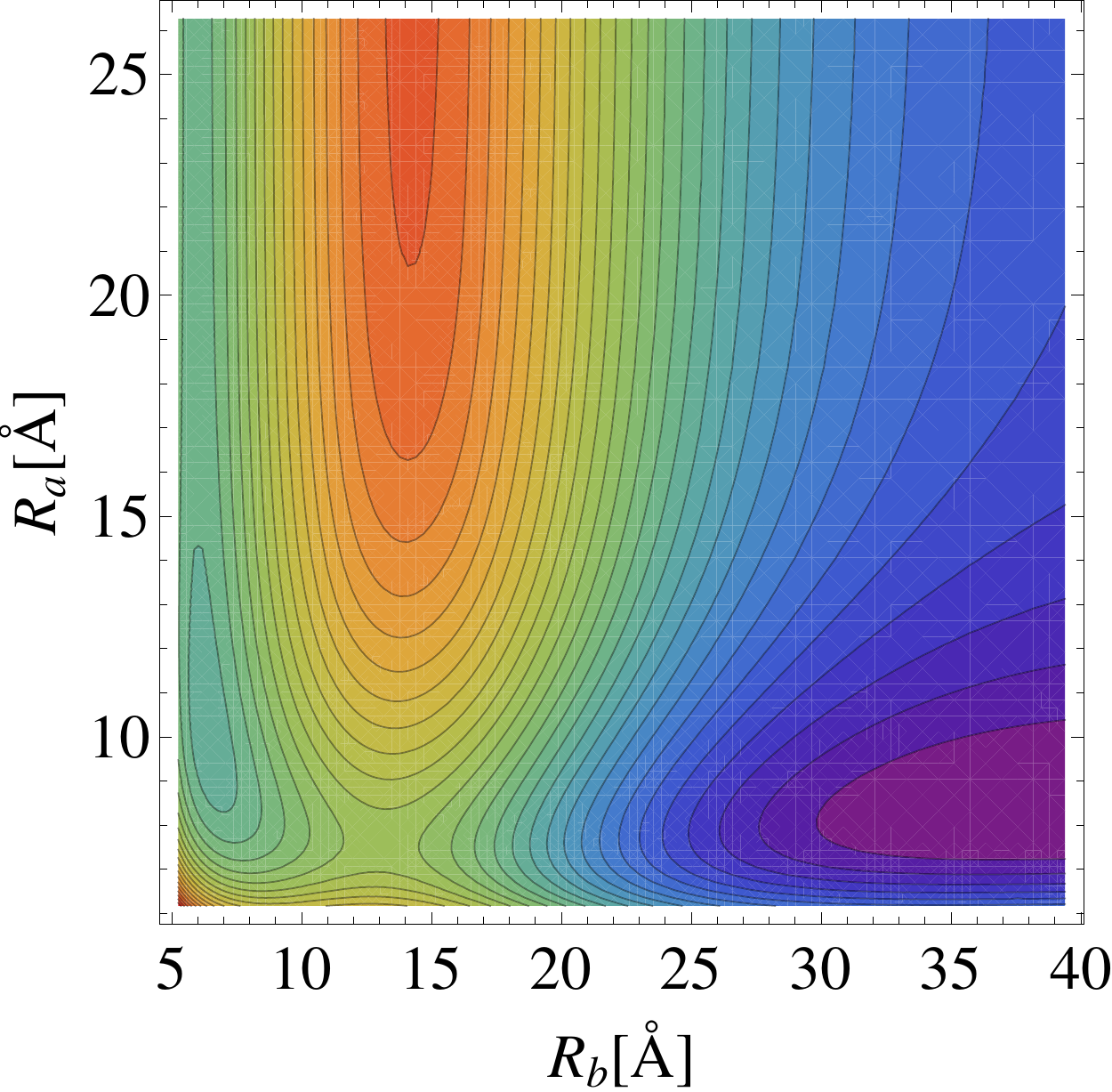}
}
\caption{The contour plot of the potential energy per dipole in the ($R_a, R_b$) -plane of regularly ordered clusters shown in Figs. \ref{fig:ldg} and  \ref{fig:3Ddg}. The third parameter in the potential energy of this ordered structure is the dipole-dipole separation distance $d$ in the $b$ direction, which has a fixed value $d$=13\AA\ obtained by minimizing the potential energy.
}
\label{fig:dchem1}
\end{figure}
\begin{figure}[tb]
\center{
\includegraphics[width=0.4\textwidth]{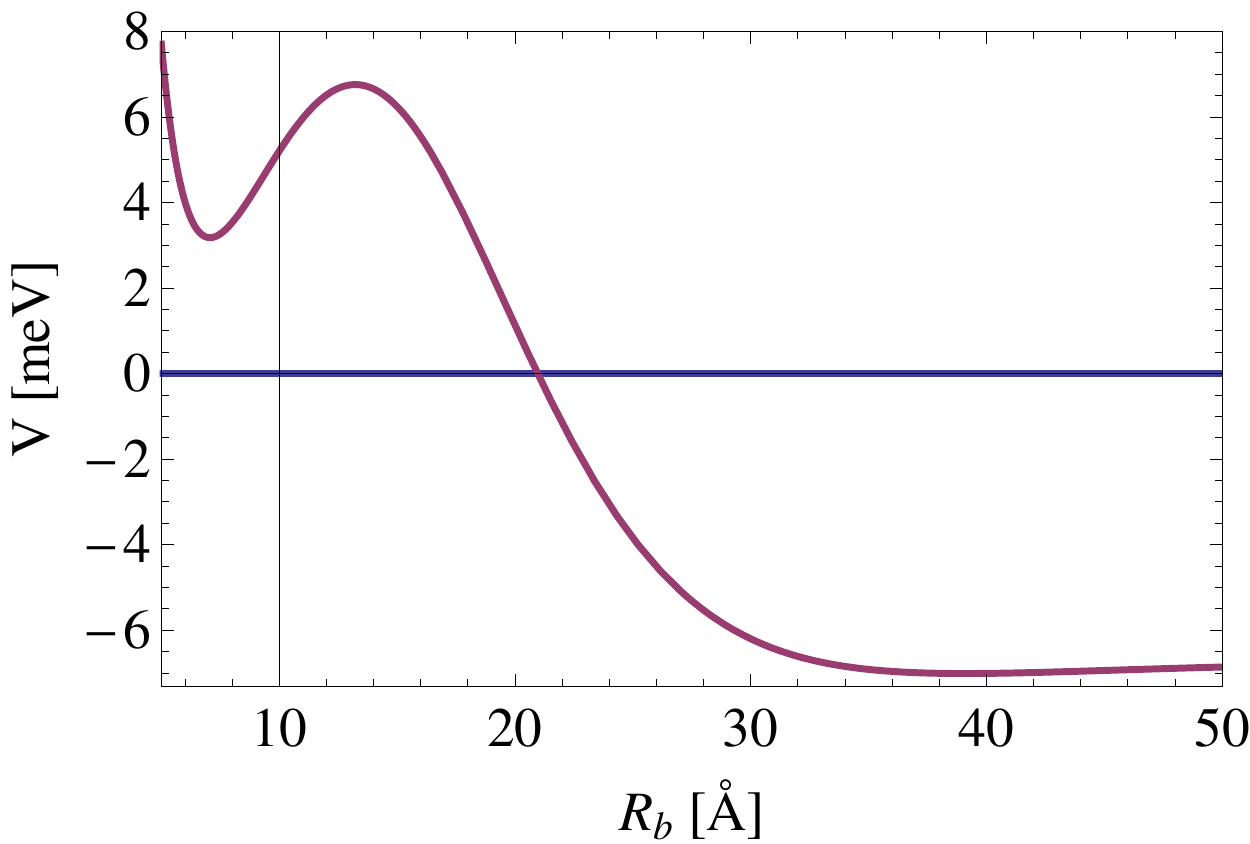}
}
\caption{The potential energy per dipole of regularly ordered clusters shown in Figs. \ref{fig:ldg} and  \ref{fig:3Ddg} as a function of separation distance between dipole-dipole pairs in $b$-direction $R_b$. Two other parameters  have fixed values, $d$=13\AA\ and $R_a$=8.2\AA. 
}
\label{fig:dchem2}
\end{figure}

In Fig. \ref{fig:dchem1} we show the contour plot of  the potential energy in ($R_a, R_b$) - plane for eight layers of 10$\times$10 dipoles in a layer organized in a regular order like in Figs.  \ref{fig:ldg} and \ref{fig:3Ddg}. The  dipole-dipole separation distance $d$=13 \AA, which gives the minimum energy in this configuration. The energy surface has local minima and the growth process may not find the absolute minimum very easily. 
Fig. \ref{fig:dchem2} then shows the potential energy per dipole as a function $R_b$, the d-d pair separation distance. We have fixed $d$=13 \AA\  as above and  $R_a$=8.2\AA. The minimum energy is -7 meV at $R_b$=39\AA. This leads to a strong directional order and the  system breaks into high and low density regions with different, nearly degenerate dipole orientations which grow slower and slower together but without the underlying, long-ranged crystal structure. Instead the system may favor the fractal like growth.

\section{Ferroelectricity at extremely  low doping}

At extremely low doping, $x\approx 0.1$\%, in La$_{2-x}$Sr$_x$CuO$_4$ and La$_2$CuO$_{4-y}$ ferroelectricity has been discovered experimentally \cite{PanagopoulosSciRep2015,Panagopoulos}. Measurements are done well inside the antiferromagnetic insulator phase and charge carrier holes are all localized. The dipolar structure between a dopant atom and clustered holes in CuO plane is strongly bound. In such dilute system distance r between dopant atoms and, consequently, dipoles is large, $r\approx 30 \AA$. The dipole-dipole interaction dominating the ordered dipolar cluster formation  diminishes like $1/r^3$ and becomes very weak. Experiments in LCO \cite{FratiniNature,PocciaNature} suggest that the growth process of macroscopic dipolar structure under radiation takes place layer by layer. Thus in a very dilute system we may consider clustering of dipoles in separated layers. As discussed above dopant ions are located in spacer layers, push down apex oxygens and bind holes in CuO plane, say below the spacer layer. The dipole points to c-direction. The interaction with a neighboring dipole pointing to the same direction is repulsive. In order to gain energy dipoles must tilt away from c-direction more than the critical angle $\theta=\pi/2-\arccos(1/\sqrt(3))$. By tilting one increases the length of a dipole and the strength of the dipole moment, but looses in binding energy.

Experiments \cite{PanagopoulosSciRep2015,Panagopoulos} show that components of polarization in c- and ab-directions are roughly equal, which is consistent with the above estimate of the tilting angle. We can calculate the dipole-dipole binding energy $q^2/(4 \pi \epsilon) d^2/r^3(1-3 \cos^2(\theta))= 1.3 meV (1-3\cos(\theta))$, where $\epsilon=21 \epsilon_0$, the length of dipole $d=3\AA$ and dipole separation distance $r=28{\AA}$, ten times the distance between Cu atoms in CuO-plane.  This gives a rough idea of the temperature range  of ferroelectric order to be below 6-7 K. 

An important condition for ferroelectricity is the two-state structure. Direction of  electric field favors parallel direction of dipoles by pushing dopant ions to one side of the spacer layer.  Switching the direction of the external electric field pushes dopant ions to the other side of the spacer layer. Bound clusters in a CuO-layer are then created to the other side of the spacer layer, direction of dipoles is reversed and polarization charges sign.  In doing this switching a typical hysteresis curve in the polarization-electric field plane should be seen. There must  be a potential barrier in moving dopant ions yielding the hysteresis behavior. Experimentally the hysteresis loop is very thin suggesting that the potential barrier is low. 

\begin{figure}[tb]
\centering
\includegraphics[width=0.5\textwidth]{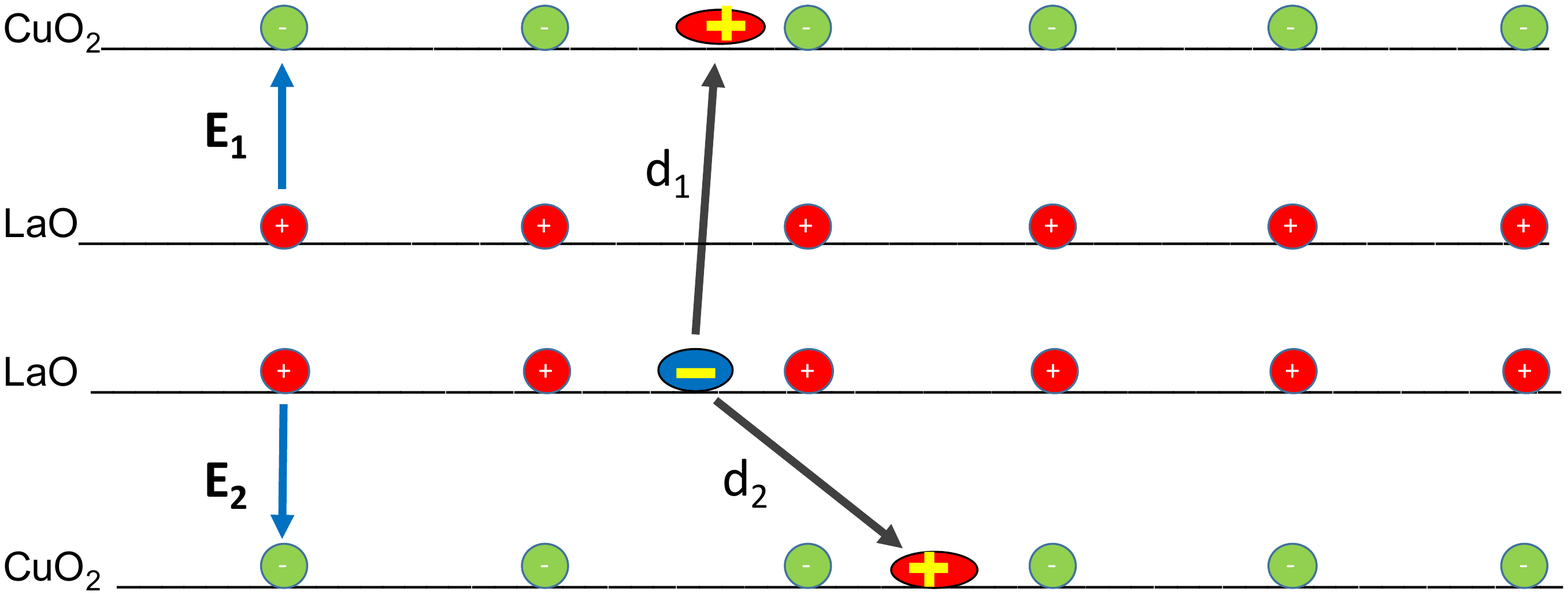}
\caption{ There is an intrinsic  electric field between the LaO layers and CuO2 layers, It is noted as  $E_1$ and $E_2$. 
 These fields form a staggered layered structure over the crystal.  The two states of the dipoles formed by a dopant atom located in the spacer layer, LaO.   The dipole may be located in the upper layer between LaO and CuO2, and its value is equal to $d_1$ or in the lower layer, below LaO where its value will be equal to $d_2$. Thus each dopant atom may be a sources of two states of a dipole.}
\label{2-state-Dipole}
\end{figure}


The energies of two dipole orientations $d_1$ and $d_2$ under the external electric fields $E_1$ and $E_2$ acting on them are $-{\bf(d_1\cdot E_1)}$ and  $-{\bf(d_2\cdot E_2)}$, see Fig. \ref{2-state-Dipole}. 
Therefore with each dopant ion we may introduce a partition function
\be
Z= \exp \beta {\bf (d_1\cdot E_1)} +\exp \beta {\bf (d_2\cdot E_2)}
\label{partition-function}
\ee
where $\beta=1/k_B T$. The polarization associated with such a dopant ion $P_i$ can be estimated with the equation
\be
P_i= d_1 \exp \beta {\bf (d_1\cdot E_1)} /Z+ d_2 \exp \beta {\bf (d_2\cdot E_2)}/Z
\label{Polarisation0}
\ee
Taking into account the staggered character of the intrinsic electrical field,  we may introduce the fields  ${\bf E_1 =E_{\rm int}+E}$ and ${\bf E_2=-E_{\rm int} +E}$.  Assuming that  ${\bf d_2=-d_1=d_i}$ we can write polarization in the following form,
\be
{\bf P_i= d_i}  \tanh(\beta  {\bf E \cdot d_i}) 
\label{Polarisation1}
\ee
In the regime of  low doping total polarization $P$ of the crystal consists of contributions from all dopants ions.  At  very low temperatures $P=n_i <P_i>$ where $n_i$ is the density of dopant ions. 

Next we  develop a  pseudo-spin formalism of this two-level system. The energy difference  $2 {\bf E \cdot d_i}=h$ corresponds to an effective field $h$.  Without an interaction between dipoles thermodynamics of our model is simply  equivalent to a standard behavior of non-interacting magnetic moments in an external field $h$. Therefore, the average "magnetization", which is here the polarization,  associated with the pseudo-spins $<{\bf d_i\cdot E}>$ is described by a standard Brillouin function for the spin $S=1/2$ as in Eq.  (\ref{Polarisation1}).

 The effective  Hamiltonian describing interacting dipoles in the staggered electric field $E(z)$, may be written in the following form:
\be
H_{eff}=\sum_{ij}  {\bf  {\cal J}_{ij}  d_i \otimes d_j} - \sum_i 
  {\bf  E \cdot d_i} 
  \label{dipole-ham}
\ee
where the second term describes the interaction with staggered electric field as discussed above. The first term of the Hamiltonian is the complex interaction between dipoles and staggered electrical field associated with the charged layers CuO$_2^{2-}$ and LaO$^+$.   It is represented  with the use of the tensor ${\cal J}_{ij}^{\alpha\beta}$, which includes also the direct dipole-dipole interaction. 
In our situations of an interacting two level system shown in Fig. \ref{2-state-Dipole}  the effective pseudo-spin  Hamiltonian becomes
\be
H_{eff}=\sum_{<ij>}  {\cal J}_{ij}  \tau^z_i \tau^z_j - h \sum_i 
\tau^z_i 
\label{pseudo-spin-ham}
\ee
here we assume that there is an interaction between nearest two level systems in the sandwich of CuO$_2^{2-}$ and LaO$^+$ pairs of layers. Note that we have introduced here the  notation $\tau^z_i \equiv {\bf d_i\cdot E}/|{\bf d_i\cdot E}|$.  One can easily show that although the coupling is a long-ranged in the first approximation we  can take into account  a nearest-neighbor interaction  which can be both repulsive, ${\cal J} >1$, i.e. an antiferromagnetic interaction or  attractive, ${\cal J} <1$, i.e. a ferromagnetic interaction
between pseudospins $\tau$. In accordance with the qualitative considerations presented above (the  state $\tau^z_1=+{1}$  corresponds to the  up dipole orientations  while  $\tau^z_1=-{1}$  is the down dipole orientation.  Longer range interactions may in general have different  sign, but usually the $nn$ interactions dominate, and this is what we will assume further on.

With this assumption we can reduce our model to an antiferromagnetic or ferromagnetic Ising model with $nn$ coupling ${\cal J}$ in a parallel field.  In this case the standard mean-field equation for the total  magnetization takes the form:

\be
\tau=<\tau>=\tanh{h -{\cal J} z\tau\over T}
\label{mean-field}
\ee
($z$ is the number of nearest neighbours), from which we can  determine the temperature dependence of $\tau$ and consequently the total polarisation of our system as ${\bf P=<d_i>} n_{dip}$.

It is convenient  to rewrite Eq. (\ref{mean-field}) as
\be
\tau=\tanh{{\tilde\Delta+{\cal J}z\left(1-\tau\right)\over T}}  ,
\label{(9)}
\ee
where $\tilde\Delta =h -{\cal J}z\tau(0)=h - {\cal J}z$ is 
the renormalized
initial $(T=0)$ splitting of these two states. If we would 
take this splitting due to $h$ to be constant
(i.e. if we ignore the second term in the argument of Eq. (\ref{(9)}),  we 
would get the conventional temperature
dependence of $\tau$ (Brillouin function)
and, consequently, of the polarisation,
which at low temperature would be exponential in temperature:
\be
\tau (T)=1- \exp \bigg( - {\tilde\Delta\over T} 
\bigg)  , 
\label{Pol-limit-zeroT}
\ee

\begin{figure}[tb]
\centering
\includegraphics[width=0.8\linewidth]{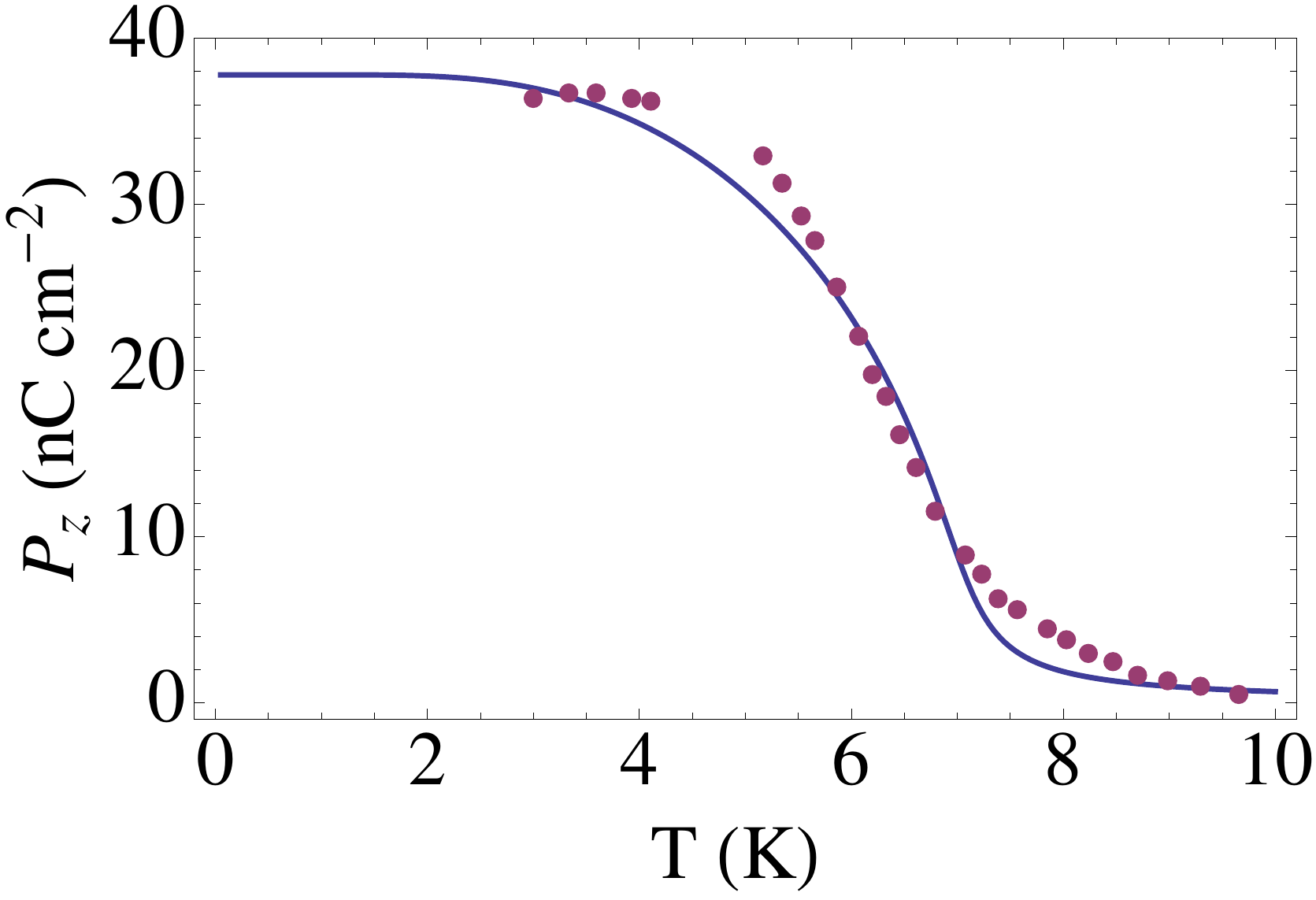}
\caption{ Out-of-plane $P_z$  electric polarization as a function of temperature $T$.   Solid line corresponds to the  theoretical calculation
performed in the framework of the developed model, when the value of dipole interaction is chosen to be ${\cal J}z=-6.9 $K and $h=0.055 $K. The dots correspond to experimental data taken from the paper\cite{PanagopoulosSciRep2015}.   }
\label{Fig2a}
\end{figure}

\begin{figure}[tb]
\centering
\includegraphics[width=0.8\linewidth]{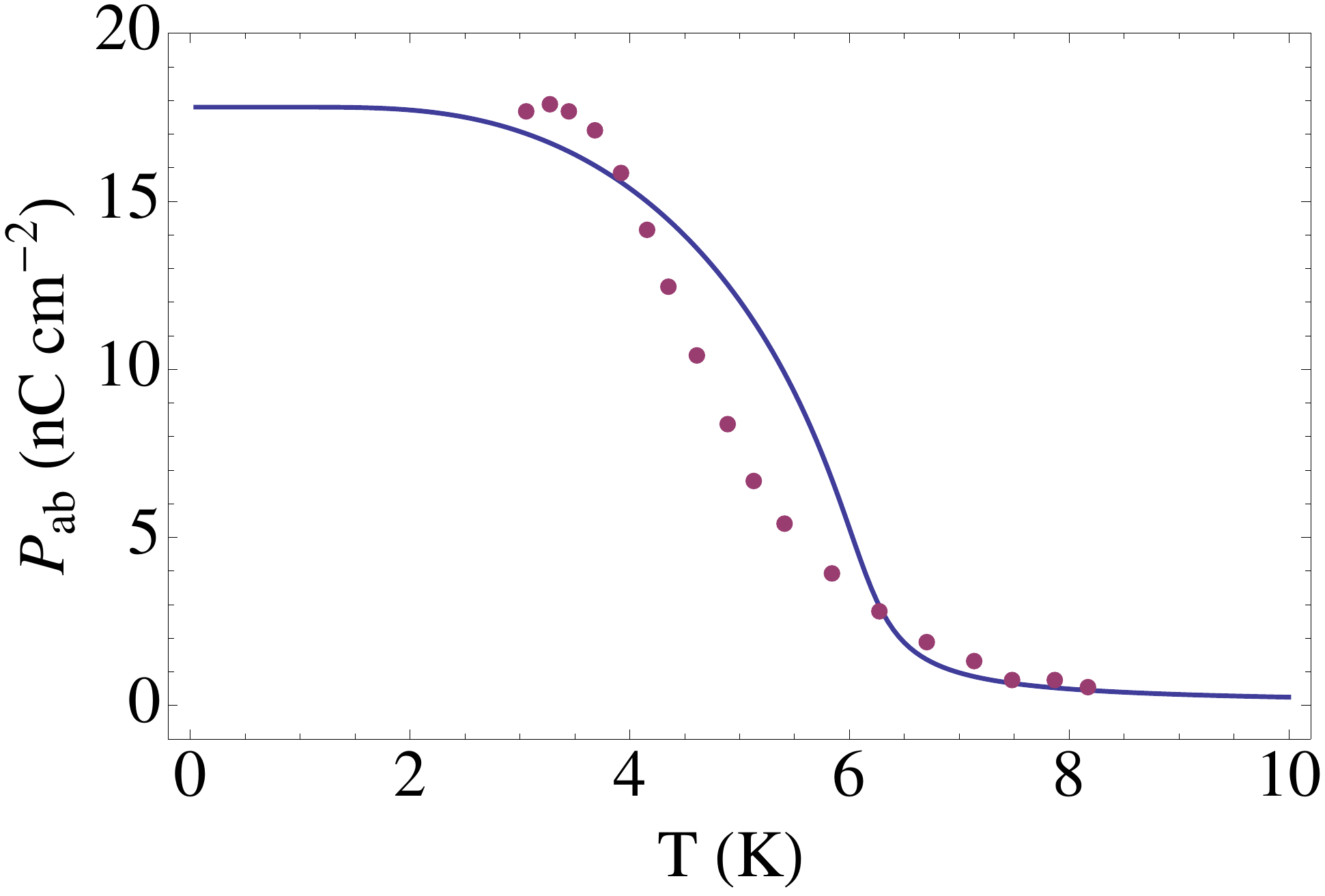}
\caption{ In-plane $P_{ab}$  electric polarization as a function of temperature.   Solid line corresponds to the  theoretical calculation
performed in the framework of the developed model, when the value of dipole interaction is chosen to be ${\cal J}z=-6.0$K and $h=0.055$K. The dots correspond to experimental data taken from the paper\cite{PanagopoulosSciRep2015}.   }
\label{Fig2b}
\end{figure}

Eq. (\ref{mean-field}) can be solved numerically. It contains two parameters $h$ and ${\cal J} z$, which can be fit to experiments. Results are shown in Figs. \ref{Fig2a} and \ref{Fig2b} where the experimental data is taken from Fig. 2 of Ref. \cite{PanagopoulosSciRep2015}. The value of $h$=0.055 K is the same in both figures, but the interaction term changes slightly from ${\cal J}z=-6.9$ K in the case of $P_z$ to ${\cal J}z=-6 $ K for $P_{ab}$.  The interaction strength  is consistent with the estimate of the dipole-dipole interaction given above. The over all strength of the polarization is fitted to the experimental value at $T=0$ to be equal $P_z(0)$=37.8  nC cm$^{-2}$ and $P_{ab}(0)$ =18.7 nC cm$^{-2}$. These values are defined by the total number of dipoles created in the samples. The figures show an excellent  agreement between the developed theory and  existing experiments. It is then clear that the critical temperature of the ferroelectric phase transition is determined by the dipole-dipole interaction at very low dopings, $x~0.1\%$. 

\begin{figure}[tb]
\centering
\includegraphics[width=0.5\textwidth]{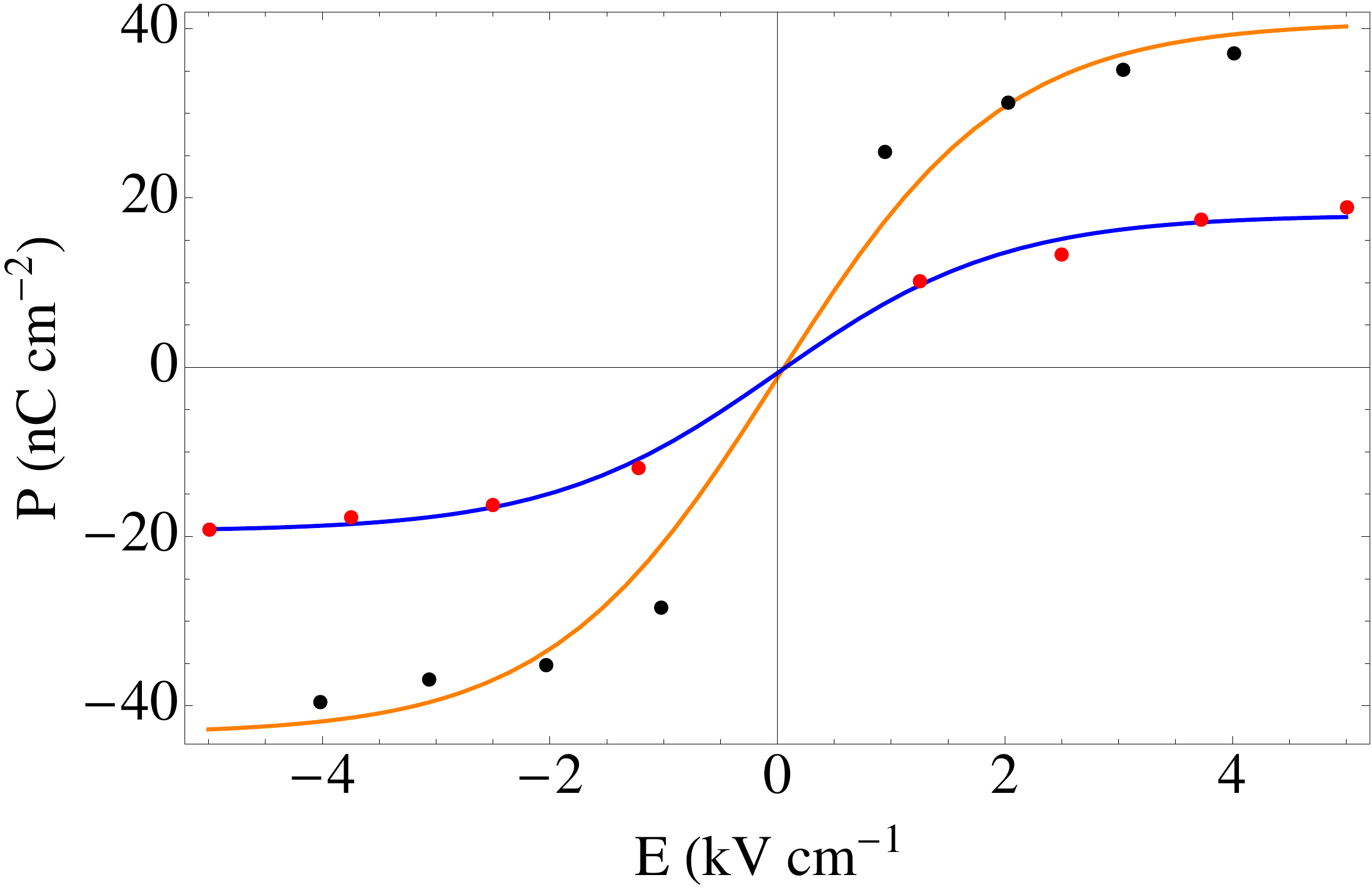}
\caption{Polarization as a function of applied electric field (P-E) curves. The data for $P_{ab}$ (red solid circles) and for $P_{c}$ (black solid circles) were measured at 2 K and presented in the Ref. \cite {PanagopoulosSciRep2015}.
 Blue and orange solid lines are calculated with the use of the derived Eq. (\ref{Polarisation1}). Here in the plot  to fit the data the saturated  $P_z$ was taken to be equal to $P_z$=42  nC cm$^{-2}$ and $P_{ab}$ =18.7  nC cm$^{-2}$
}
\label{P-E-plot}
\end{figure}

In Fig.  \ref{P-E-plot} we show the behavior of the strength of polarization $P$ as a function of applied external field $E$ at the temperature $T=2K$. Again we have a very good fit of our theory to the experiments of Ref. \cite{PanagopoulosSciRep2015}  for for the  LCSO samples
with very small doping $x=0.1\%$. Polarization along the c-direction dominates due to the staggered field.
Note also that the dependence of the in-plane polarization $P{ab}$ on applied electric field $E$, see the blue curve (the  developed theory) and red dots (the experimental data from the Ref. \cite{PanagopoulosSciRep2015} are well described  by a standard Brillouin function for the spin $S=1/2$, which coincides with the derived Eq. (\ref{Polarisation1}). However the similar dependence for the  z-component of the polarisation, $P_z(E)$ shows already small deviation between the theory and experiments, that indicates that the dipole-dipole interaction should be here properly taken into account.

\section{The resonance plaquettes at optimal doping are viewed as Planckian dissipators}

For free hole current carriers each dipole is a source of strong scattering. The scattering becomes stronger when the doping increases and reaches its maximum at the optimal doping. 
To estimate the scattering time $\tau$ we have to look  into the detailed structure of individual dipole clusters and their evolution with doping. 
Here each dipole is formed by a cluster of four holes  located in the CuO plane and created by the many-body over-screening  of the Coulomb attraction of these holes to the impurity or electron polarons located in the LaO spacer layer. 
The whole process of the cluster formation is accompanied by the squeezing of the neighbouring oxygen octahedron and creation of heavy polarons located in LaO spacer. 
That process of  the oxygen octahedron  squeezing is known as anti-Jahn-Teller effect and in turn it may provide four holes for the cluster and the singlet electron pair for the flat band as shown in Fig. \ref{fig:LaCuO}. 
The rest doped free Zhang-Rice holes form a  strongly-correlated quantum liquid.  The many-body interaction of these holes in the strongly-correlated regime provides the over-screening of the charged impurities  and polarons located in the LaO layer  leads to the cluster formation. 
It is important to note that even without impurities the fluctuation  associated with the squeezing of the  oxygen octahedron locally, originated in such strongly correlated liquid may lead to formation of a similar cluster of four holes bound to polarons.  However, such states are metastable and their structure depends on the doping level.

With increasing temperature clusters become metastable. The life time of such a fluctuating cluster, $\tau_{life}\sim \hbar/2E_{bin}(x)$,  depends on its binding energy at a given doping $x$.  It is maximal near optimal doping where $x\sim0.2$. This effectively means that in the region of the optimal doping there are continuous quantum fluctuations associated with formation of these resonance plaquettes consiting of four holes. In other words they are decoupled from dopant impurities and become mobile. 

Four holes clusters behave as quasi-classical objects moving along in a nearly flat band. Obviously such continuous fluctuations play huge role in the scattering of the remaining still very mobile free holes. As classical objects the thermal energy of clusters is $ E_{kin}\sim k_B T$.  The motion of free current carriers through such a clustered media is very dissipative. Free holes can be individually trapped by cluster fluctuations  and turn all their energy irreversibly into heat. Using the Heisenberg uncertainty principle we can estimate the scattering time of free holes as the time they leave in the cluster fluctuation, $ E_{kin} \tau \sim  k_B T \tau\geq \hbar/2$.   Such a process takes a characteristic (relaxation) time, 
\be
\tau=\tau_{Planck}=\hbar/2k_B T \,,
\label{Planck}
\ee
which is the shortest possible relaxation time  named as the Planck time\cite{Zaanen_2004}.

\section{Quantum Criticality and comparison with experiments}

The resistivity of the mixture of two components -- free current carrier holes and  four hole clusters -- can be described by Drude formula, $R=m / n_h(x) e^2 \tau$, where $m$ is the effective mass of holes and $n_h(x)$ is their density as a function of doping $x$. Near the optimal doping $x\sim0.2$ the binding energy of clusters vanishes and fluctuation in the cluster formation will dominate the motion and transport of charge carriers. Inserting the scattering time of  Eq. (\ref{Planck}) into the Drude formula we get the linear T-dependent contribution to resistivity,
\be
R=\frac{2m k_B T}{n_h e^2 \hbar}=\alpha_1(x) T\,.
\label{resistivity}
\ee
In the underdoped region the binding energy of clusters increases with decreasing doping or increasing $r_s$ as shown in Fig. \ref{activation}.  Trapping of free holes into clusters becomes less probable and their transport becomes more Fermi liquid like with $T^2$ term in the resistivity. 

 We are now in position to make a comparison with experiments on LSCO \cite{Cooper2009,PNASBarisic}, where in the region of optimal doping the T- linear dependence of resistivity has been in detail investigated. The value of the effective mass $m=6.7m_e$ and the hole density $n_h(x)$  are determined above from the comparison with the Hall coefficient measurements \cite{PhysRevB.75.024515-Ono}. 
Inserting these numbers into Eq. (\ref {resistivity}) we get the T-linear coefficient. At $x=0.18$ $\alpha_1=1.14 \mu \Omega $ cm/K and at $x=0.21$ $\alpha_1=0.82 \mu \Omega $ cm/K. These values are within the error bar of the experimental results\cite{Cooper2009}, which is surprising for such an order of magnitude estimate. When $x<0.18$ the increasing binding energy of clusters diminishes  the T-linear behavior and increses the normal Fermi liquid contribution as also seen in experiments \cite{Cooper2009}.

 These results confirm that the above theoretical calculations of the pseudogap  (see, Fig.2)  associated with the formation of the four holes clusters - the resonance plaquettes provide not only qualitative but also quantitative description of the underdoped and optimally doped cuprates, That also anticipates formation of anti-ferroelectric fractal structures  which are all melted at optimally doping where a quantum phase transition arises.

\section{Discussion and Conclusions}
We have shown that in LSCO and other related compounds Coulomb interaction between charge carriers and  doped ions is very important. The strong electronic correlations tightly connected with anti- Jahn-Teller lattice distortions lead to   bound states of  four holes forming a resonance plaquette.  The formation of the plaquettes located in the CuO plane is associated with the over-screening of the Coulomb interaction. In the underdoped region  they are bound with dopant  atoms creating a  cluster which has a dipole moment due to the separation of the plaquette and dopant impurity having opposite charges. 

Recently Poccia et al. \cite{Poccia25092012} proposed a mysterious second growth mechanism, which they have studied at T=85K. They relate that to the dynamics of local lattice distortion in the CuO plane. Also, in that case the scale free clusters, Q3, appear. The key question is what happens when the lattice is distorted and how does this distortion arise?  The answer should be connected with the continuous X-ray illumination, which is a necessary condition for a growth of these snowflakes. 

Indeed such an irradiation may create excitons of intermediate radius. In the Ref. \cite{FeoRashba} it was shown that if  the excitons of intermediate radius are created they may create pairs of lattice defects as interstitial and vacancy. Moreover such excitons may have a radiationless decay into the pair consisting of the negatively charged oxygen interstitials, i-O, and the positively charged vacancy, V+, which may be, in general, decoupled. Such excitonic mechanism of the local lattice defect formation and distortions has been observed experimentally in Refs. \cite{Fugol1988,Savchenko1985}, indeed. When such a vacancy will approach to another i-O, they may annihilate. The creation of such virtual or short time living pairs increases significantly the mobility of the i-O's and form as special pattern of the local lattice distortions, noted as Q3 in the Ref. \cite{PocciaJSNM,Poccia25092012}.  The pairs located in the LaO plane form dipoles lying in the plane that change the structure of the clusters. In fact there due to the presence of these extra coupled and decoupled i-O-V pairs the mobility of i-O's increase. This leads to a faster formation of the new type of clusters different from the ones described in the Refs. \cite{FratiniNature,PocciaNature,LittlewoodNature}. The additional dipoles associated with the new pairs created with X-ray illumination lead to a new period of the stripe ordering in these new type of clusters noted as Q3 in the Ref. \cite{FeoRashba}.  Besides the excitons generated the LLD-Q3 may arise only in the deeply underdoped case as excitons at all. The probability to such excitons with X-ray illumination very fast vanishes with the doping. This explains why the Q3 clusters are arising only in the underdoped cuprates.

Possibly, also the oxygen atoms may be displaced from their original positions in tetragons. They take one or two electrons with them, act as ions and leave behind holes, which are responsible of the superconducting state when the temperature is lowered below the critical temperature.  These ions are not lifted into the LaO layer and are more mobile. However, negative ions form again bound states with holes and dipoles orient in the growth process as in the case of i-O and form fractal clusters. The only difference between these two mechanisms is the location of the oxygen ions and shows up as slightly different stripe order of the nano grains. 

Recently,  the nature of oxygen dopant-induced states found at $-0.8eV$ in  Bi$_2$Sr$_2$CuO$_{8+x}$ by measuring ARPES spectra in a wide photon energy range has been investigated.\cite{PhysRevB.74.094512-Richard} The found resonance profile of the corresponding nondispersive peak indicates an unexpected mixing with Cu states.  The A$_{1g}$ symmetry of the peak 
suggests that the oxygen dopant-induced states are mixed with Cu through the O$_{dopant}$-O$_{apex}$ 2 p$_z$-Cu~ 3 d$_{3z^2-r^2}$ channel. \cite{PhysRevB.74.094512-Richard}

Experiments indicate that the doped cuprates have a mixture of characters. To see this, we start with an undoped parent compound such as La$_2$CuO$_4$. This material is an insulator with a charge-transfer gap of $\sim$ 2eV. Antiferromagnetism develops within the CuO$_2$ planes as a consequence of strong onsite Coulomb repulsion between electrons in the same Cu 3d $x^2-y^2$ orbital. The effective magnetic interaction is well characterized by the superexchange mechanism, and the magnetic excitation spectrum is described quite well by spin-wave theory with nearest-neighbor superexchange energy.\cite{Tranquada20121771}

Thus, the dipolar clusters appear when dopant oxygen ions in LaO layers trap charge carriers in CuO planes. The density of dipole moments increases with doping and vanishes at the pseudogap temperature. The binding energy of holes into these dipolar clusters defines the pseudogap, which we identify also as the activation energy of the two fluid model used by Gorkov and Teitelbaum. \cite{PhysRevLett.97.247003-Gorkov,PhysRevB.75.024515-Ono,RevModPhys.82.1719-Phillips} Dipole moments have a strong interaction with  electromagnetic radiation and subjected to such radiation  dipoles become mobile. Due to the dipole-dipole attraction existing for some  dipole orientations and repulsion for other orientations they may form clusters. This finding explains the recent observation of the self-organization of mobile oxygen dopant ions in LCO  by Bianconi et al. \cite{FratiniNature,PocciaNature}.  Under irradiation oxygen ions resting in the LaO layer as well as dipoles get excited and are forced to move towards the energetically most favourable positions. 

 Ferroelectricity in La$_2$CuO$_{4+y}$ and La$_{2-x}$Sr$_x$CuO$_4$ was reported at exceptionally low doping.\cite{PanagopoulosSciRep2015,Panagopoulos}. We have shown that the dipole formation and its reduction to the two-level system under external electric explains the behavior of the  ferroelectric polarization  as a function of temperature and  electric field. It could also be responsible of the peculiar behavior of the dielectric  constant  \cite{PhysRevB.72.064513_Wang}. The resonance plaquettes  described in this paper may play a very important role in the mechanism of high temperature superconductivity. Their presence in the vicinity of the hole band edge may lead to shape resonances as described recently by Bianconi\cite{Nature.Physics.Bianconi2013}, that can be an origin of the Cooper pairing in cuprates. Our results strongly suggest that many phenomena in cuprates such as pseudogap formation and high temperature superconductivity are interaction driven and therefore these materials belong to a new class of holographic superconductors\cite{Nature.Physics.Hartnoll2013}. 
  	
\section{Acknowledgements}
We thank Antonio Bianconi, Lev Gorkov,  Danya Khomskii,  Christos Panagopoulos, Montu Saxena, Grisha Teitelbaum and Jan Zaanen for useful discussions. We also thank NanoSc-Cost Action MP1201 for financial support.

\bibliographystyle{elsarticle-num} 

\end{document}